# *In Situ* Dynamics of the Microscopic Crystallographic Dehydration Pathway in a Model Channel Hydrate, Theophylline

N. Koniuch[1,2*], S. T. Pham[1,2], M. Danaie[3], F. Costa[2], Z. Aslam[2], S. Foster[1,2], H. Blade[4], L. Hughes[4], N. Hondow[1,2], R. Drummond-Brydson[1,2], S. M. Collins[1,2,5,6*] and A. P. Brown[1,2*]

[1]*School of Chemical and Process Engineering, University of Leeds, UK*

[2]*Bragg Centre for Materials Research, University of Leeds, UK*

[3]*Electron Physical Science Imaging Centre (ePSIC), Diamond Light Source, Didcot, UK*

[4]*AstraZeneca, Oral Product Development, Pharmaceutical Technology & Development, Operations, Macclesfield, UK*

[5]*School of Chemistry, University of Leeds, UK*

[6]*Department of Materials, Royal School of Mines, Imperial College London, UK*

**Email: N.A.Koniuch@leeds.ac.uk, S.M.Collins@imperial.ac.uk, A.P.Brown@leeds.ac.uk*

## Graphical abstract

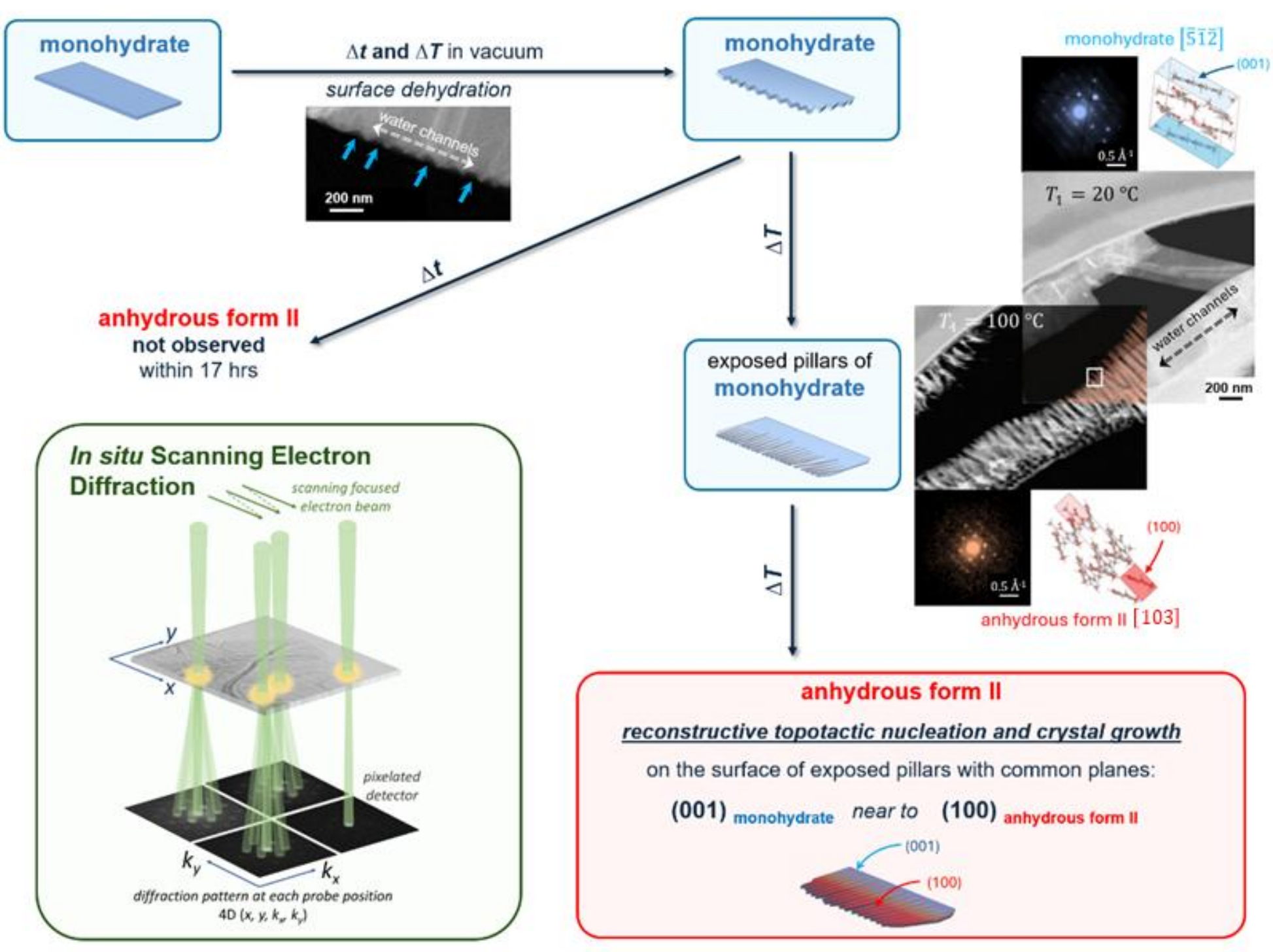

## Abstract

Solid-state phase transformations in molecular crystal hydrates govern stability and functional performance across a range of applications, including pharmaceutical, agrochemical and coordination framework materials. During dehydration, these hydrates can undergo substantial structural reorganisation involving changes in molecular orientation, intermolecular interactions, and lattice symmetry. Despite extensive study, the microscopic crystallographic pathways by which such transformations proceed remain poorly understood. Here, we investigate the dynamics of solid-state dehydration of theophylline monohydrate as a model molecular hydrate using *in situ* low-dose scanning electron diffraction (SED). Simultaneous observations of changes in morphology and crystallographic phase and orientation mapped across single particles reveal how complete dehydration proceeds via a two-step, reconstructive topotactic solid-state transformation: anisotropic, surface-specific mass loss of material near water channel sides (suggesting the monohydrate adopts a non-centrosymmetric crystal structure) is followed by surface-localised nucleation and growth of anhydrous form II on the parent monohydrate while preserving similar molecular orientations at a common plane. By providing direct, local crystallographic insight into hydrate dehydration, this work demonstrates how surface-controlled mass loss, morphological changes, and lattice orientation jointly govern solid-state transformations in molecular hydrates. More broadly, it establishes low-dose SED as an effective approach for probing dynamic phase transformations in beam-sensitive molecular crystals.

## Introduction

Solid-state phase transformations govern the physical properties of crystalline materials across metals (1, 2), ceramics (3), inorganic solids (4), and molecular crystals (5, 6). For example, in steels, phase transformation to martensite or bainite can determine mechanical strength, toughness, and functional behaviour, and have been investigated using *in situ* and *ex situ* electron microscopy to resolve nucleation, interface motion, and crystallographic relationships during transformation (7-9). These studies demonstrate how direct structural observation has enabled detailed understanding of transformation pathways in atomic solids. In molecular crystals, solid-state phase transformations are similarly important, particularly in pharmaceutical and agrochemical materials where a specific polymorph can control solubility, dissolution rate, and bioavailability (10, 11). Yet, despite extensive experimental and theoretical work, the microscopic pathways of solid-solid phase transformations in molecular crystals remain less well established. In contrast to simple atomic solids, molecular rearrangement in the solid-state is constrained by the size and shape of the molecules in combination with the strength and direction of the intermolecular bonds, as well as the presence of disorder at grain boundaries and defects. This complicates the direct application of classical nucleation-and-growth models developed for metals and inorganic materials (12, 13).

Classical descriptions of first-order solid-state phase transformations assume nucleation at pre-existing defects followed by interface-controlled growth, with structural rearrangement occurring locally at a moving phase boundary. While this framework has been successful for many atomic solids, experimental studies on molecular crystals increasingly indicate that transformations may involve collective or partially cooperative molecular motion over finite length scales, particularly when parent and daughter phases share similar packing motifs or crystallographic planes (12, 14). Crystallographic analyses further show that solid-solid transformations can preserve specific orientation relationships between phases even when substantial reconstructive rearrangement occurs. Such behaviour is commonly described as topotactic, or quasi-topotactic when orientation preservation is incomplete (15). Despite these advances, the extent to which molecular solid-state transformations proceed via strictly local rearrangement versus cooperative motion, and how the balance depends on crystallographic similarity between parent and daughter phases, hydrogen-bond topology, and anisotropic strain accommodation, remains unresolved (13, 14).

Resolving these mechanistic questions is particularly important for pharmaceutical and agrochemical molecular solids, where hydrate formation and dehydration commonly occur during crystallisation, wet granulation, aqueous coating, and storage under humid conditions and can result in substantial changes in product performance (16-18). Understanding this behaviour is especially critical for channel-type hydrates, where water occupies anisotropic tunnels that enable rapid hydration and dehydration (19). Such systems are prone to forming metastable intermediates or structurally disordered states during water loss, complicating the control of processing conditions and prediction of solid-state stability (20). Addressing these limitations requires the development of experimental approaches capable of resolving local structural changes and crystallographic relationships within the solid during transformation.

Electron microscopy offers the spatial resolution required to probe nanometre-scale structural changes within individual crystals, but its application to organic molecular solids is challenging due to a pronounced sensitivity to electron beams (21-23). As a result, conventional transmission electron microscopy (TEM) imaging often leads to rapid loss of crystallinity, limiting achievable resolution and, in some cases, the feasibility of structural analysis (21, 22, 24). Selected-area electron diffraction (SAED), whilst providing a single diffraction pattern from a fixed region, typically several hundred nanometres across, cannot resolve finer spatial heterogeneity or early structural changes within a phase. Although high-resolution TEM imaging of organic crystals is possible under carefully optimised low-dose conditions (25, 26), rapid beam-induced degradation of the crystalline structure usually limits the number of images that can be acquired from the same region, making *in situ* image series acquisition experimentally challenging.

Low-dose scanning electron diffraction (SED) overcomes these limitations by rastering a low-current electron probe of a few nanometres in diameter (formed with a small convergence angle to produce a so-called 'pencil beam') across the specimen while recording a two-dimensional diffraction pattern at each pixel (Fig. 1a) (27, 28). This approach provides nanometre-scale spatial (i.e. real-space) resolution together with reciprocal-space resolution at every probe position sufficient for large molecular crystal unit cells, enabling direct mapping of local structural changes, orientation differences, and nanoscale domains that cannot be accessed by SAED or bulk diffraction methods. SED has demonstrated broad applicability for analysing beam-sensitive materials, such as: lamellar and nematic microstructures in waxes (29), framework topology and defect chemistry in metal–organic frameworks (30), crystalline domains and orientation in polymers and co-polymers (31, 32), planar defects in molecular crystals such as xanthine derivatives (33) and even crystallographic dislocations and Burgers vector identification in molecular crystals (34). SED can operate at very low total electron fluences (just a few electrons per $Å^2$), enabling repeated scanning of the same region and thus offers the potential for *in situ* or dynamic observation of local phase changes under thermal or environmental stimuli (35-37), something that remains largely unexplored for molecular crystals during dehydration.

Theophylline, a methylxanthine derivative widely used as a bronchodilator, has long served as a model system for studying channel hydrate-anhydrate transformations (38-43). The anhydrous form II ($C_7H_8N_4O_2$) is the thermodynamically stable polymorph under dry, ambient conditions and is used as an active pharmaceutical ingredient (API). In contrast, theophylline monohydrate ($C_7H_8N_4O_2 \cdot H_2O$) readily forms during wet granulation or upon exposure to humidity. While hydration of anhydrous form II to the monohydrate occurs in a single step (41), dehydration can proceed either (i) directly via a single step (39) or (ii) via a metastable intermediate (form III) before conversion to anhydrous form II (Fig. 1b) (43, 44). Form III has been described as a 'dehydrated hydrate', retaining the stacked dimer framework of the monohydrate, but lacking water in the channels and exhibiting increased structural disorder (45). The transformation pathway is governed by the dehydration rate, which is controlled by temperature and relative humidity: slower dehydration at lower temperature and/or higher humidity favours formation of the metastable intermediate, whereas more rapid dehydration promotes direct conversion to the anhydrous form II (44, 46, 47). The presence of metastable form III in a pharmaceutical product can lead to tablet hardening and altered dissolution performance, resulting in batch-to-batch variability and bioavailability issues (47, 48). Despite extensive study, the structural pathways by which these transformations proceed remain incompletely understood.

Theophylline monohydrate contains hydrogen-bonded water channels that zigzag along the *a*-axis that stabilise a monoclinic lattice; both a $P2_1/n$ space group and its non-centrosymmetric subgroup $P2_1$ have been reported, the latter structure reflecting polarity within the water-chain hydrogen-bond network (Fig. 1c, Table S 1, S 2) (49, 50). In contrast, anhydrous form II crystallises in an orthorhombic structure with denser packing ($Pna2_1$) (Fig. 1c, Table S 1) (51). Dehydration, therefore, involves a change in crystal symmetry from monoclinic to orthorhombic, including substantial molecular reorientation within hydrogen-bonded layers and the collapse of the water-stabilised channels. A purely (solid-state) displacive mechanism would require large molecular shifts, whereas a dissolution-reprecipitation pathway would be expected to randomise lattice orientation and eliminate crystallographic registry between the phases. Nevertheless, bulk dehydration experiments frequently report rapid transformation with retention of crystalline order, implying that extensive molecular rearrangement can occur directly in the solid-state. Several mechanistic models have been proposed, including Avrami-Erofeev nucleation and growth kinetics (43, 47, 52), Fickian diffusion-based descriptions (53), and phase-boundary-controlled mechanisms (52). However, these models do not explain how molecular rearrangement propagates at the nanoscale or how crystallographic registry can be preserved during such a substantial structural reorganisation.

Here, we address this challenge by examining the dehydration and phase transformation of theophylline monohydrate under two sets of conditions with different transformation rates: *in situ* vacuum dehydration ($\Delta t$) and controlled, rapid *in situ* vacuum heating ($\Delta T$). Low-dose SED is used to map phase changes and crystallographic orientation at the single-particle level (Fig. 1d). By combining virtual dark-field imaging with diffraction analysis at nanometre spatial resolution before, during, and after dehydration, we can directly resolve phase coexistence, orientation relationships, and local progression of the transformation. This approach provides new structural insight into the dynamics of solid-state dehydration of a molecular crystal and demonstrates the broader potential of *in situ* low-dose SED for studying dynamic phase transformations in other molecular crystals.

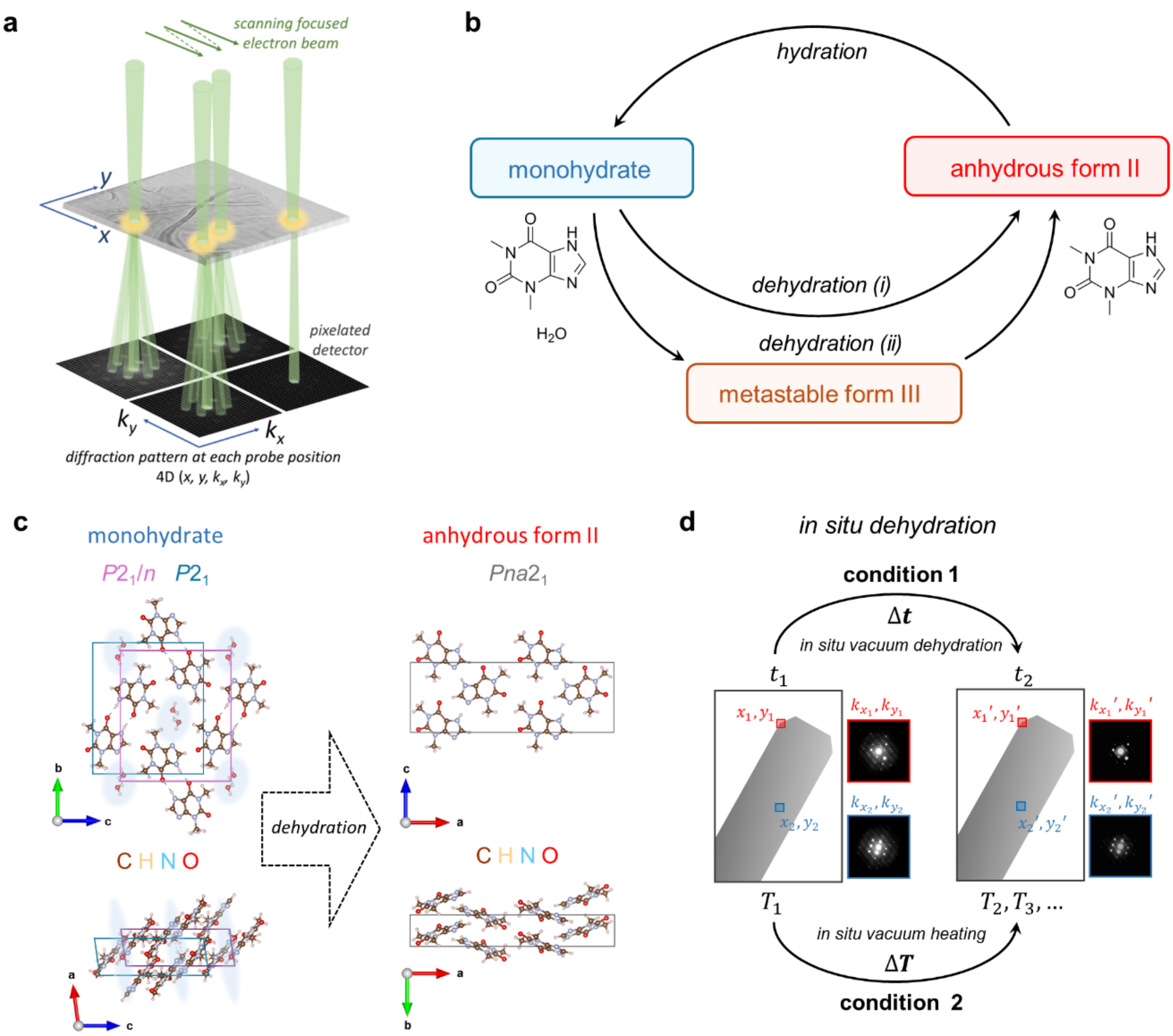

*Fig. 1. The in situ low-dose SED methodology and overview of theophylline phase transformations.* ***a*** *Schematic representation of the SED methodology.* ***b*** *Schematic of phase transformation pathways between theophylline monohydrate and anhydrous form II. Hydration occurs as a one-step process, whereas dehydration can proceed through either one or two steps depending on conditions (temperature and/or relative humidity).* ***c*** *Molecular projections of the two possible unit cells of theophylline monohydrate and of anhydrous form II. CCDC 183780 (monohydrate,* $P2_1/n$*) (50), CCDC 2264675 (monohydrate,* $P2_1$*) (49) and CCDC 128707 (anhydrous form II) (51).* ***d*** *In situ SED approach for investigating dehydration under two experimental conditions: (1) in situ vacuum dehydration (*$\Delta t$*) and (2) in situ vacuum heating (*$\Delta T$*).*

## Materials and Methods

### *Sample preparation*

Theophylline anhydrous form II ($C_7H_8N_4O_2$) was purchased from Sigma-Aldrich (CAS 58-55-9; ≥ 99% purity). Theophylline monohydrate ($C_7H_8N_4O_2 \cdot H_2O$) was obtained by evaporative recrystallisation from ethanol-water mixtures. Previous studies have shown that the monohydrate can be prepared from both aqueous and mixed organic-aqueous solvents (44, 54, 55). The addition of an organic component was reported to reduce solubility and increase supersaturation, enabling crystallisation at lower temperatures (56). Anhydrous form II was dissolved in ethanol:water (3:2 molar ratio). The solution was covered with aluminium foil and heated to approximately 55 °C on a magnetic stirrer with continuous stirring. The water activity ($a_w$) of this mixture was estimated to exceed 0.67 at 25 °C (57) and likely increased during heating due to ethanol volatility. The solution was cooled gradually to ~35 °C under continuous stirring over 3 h. Upon initial precipitation, a drop was deposited either onto holey carbon TEM grids (EM Resolutions) for initial characterisation by TEM and *in situ* vacuum dehydration experiments or onto DENSsolutions Wildfire Nano-Chips with amorphous $Si_3N_4$ support films for *in situ* vacuum heating studies. In both cases, the drop was left to evaporate before loading into the microscope vacuum. Slow evaporation yielded needle-shaped crystals suitable for electron microscopy.

All monohydrate samples were stored in grid boxes within a desiccator at 93% relative humidity, maintained using a saturated potassium nitrate ($KNO_3$) solution, to preserve the hydrate phase. The hydrate phase remained stable when sealed at room temperature for approximately four days. After complete precipitation, the solids were collected by filtration, surface dried at ambient conditions, and characterised by powder X-ray diffraction (pXRD).

### *Powder X-ray Diffraction*

The pXRD pattern of recrystallised theophylline monohydrate was collected using a Bruker D8 Advance diffractometer with Cu $K\alpha$ radiation (λ = 1.54180 Å). To minimise dehydration, data were collected at a scanning speed of 0.5 s per 0.02° step over a 2θ range of 5° to 40°. The powder sample was gently pressed into the holder with a glass slide to ensure coplanarity between the powder and the sample holder.

Using TOPAS Academic v.7 (58), a Rietveld refinement (59) was performed employing the fundamental-parameter approach (60, 61) for peak-profile modelling and Chebyshev polynomials to describe the background. Preferred orientation effects were accounted for using spherical harmonics. Three different crystal structures of theophylline monohydrate were found in the Cambridge Crystallographic Data Centre (CCDC) and were initially considered for the Rietveld refinement: THEOPH ($P2_1$, non-centrosymmetric packing; CCDC 1270454; (62)), THEOPH01 ($P2_1/n$, centrosymmetric packing; CCDC 183780; (50)) and THEOPH08 ($P2_1$, non-centrosymmetric packing; CCDC 2264675; (49)). However, attempts to fit the THEOPH crystal structure to the experimental data confirmed that it does not describe the theophylline monohydrate sample very well. Also, the CCDC reports an error in the description of this structure, which is associated with the coordinates and symmetry of the N7, C8 and hydrogen atoms in the theophylline molecule, as well as the H10 atom of the water molecule, all leading to an incorrect void space in the structure. Consequently, THEOPH01 and THEOPH08 crystal structures were used here as structural models to fit the initial pXRD.

### *Electron microscopy*

Initial TEM characterisation was performed on an FEI Titan$^3$ Themis at the University of Leeds operated at 300 kV and equipped with a continuously variable gun lens (as part of its monochromator assembly) used here for electron dose control. Bright-field imaging (BF-TEM) and SAED patterns were acquired under low-dose conditions, using an electron flux of 0.08-0.09 $e^{-}$ Å$^{-2}$ $s^{-1}$ calibrated with a Faraday cup. Exposure times were typically 2 s for imaging and 4 s for diffraction acquisition, with several additional short low-dose exposures used for region selection and focusing.

Low-dose SED data were collected on a JEOL ARM300CF operated at 300 kV with a cold field emission gun at the electron Physical Sciences Imaging Centre (ePSIC), Diamond Light Source (UK). The microscope was configured for nanobeam diffraction as reported previously (29, 34) using a 10 μm condenser aperture to produce a convergence semi-angle <1 mrad, corresponding to a diffraction-limited probe of ~5 nm diameter. Briefly, the probe-forming aberration corrector elements were switched off, and a low-convergence angle probe was formed by adjusting the condenser optics. The probe current was set to 1 pA (measured with a Faraday cup), and the exposure time was 1 ms per probe position. These conditions gave an estimated electron fluence of ~8 $e^{-}$ Å$^{-2}$ at each probe position (per diffraction pattern). Low-dose TEM studies on anhydrous theophylline report a critical fluence of 36 ± 8 $e^{-}$ Å$^{-2}$ at 300 keV (63). Diffraction patterns were recorded at each probe position using a Merlin-Medipix hybrid counting direct electron detector (Quantum Detectors, UK). The beam was rastered with a step size of 7.8 nm over a $256 \times 256$ pixel scan, resulting in nominally non-overlapping probe positions. The final scan column was removed due to fly-back overexposure, and one row was cropped to maintain a square $255 \times 255$ dataset.

Preliminary morphological characterisation and *in situ* TEM experiments were performed using a Gatan double-tilt holder and a Gatan single-tilt *in situ* heating holder (model 628) using the University of Leeds FEI Titan$^3$ Themis. A JEOL single-tilt holder and a DENSsolutions Wildfire heating holder were used for *in situ* SED experiments at ePSIC. For *in situ* vacuum dehydration, regions of interest were scanned immediately after loading and then again after ~17 h in the TEM vacuum, during which time the sample was kept in the TEM column. For *in situ* thermal dehydration, SED data were acquired from selected regions at room temperature, 65 °C, 85 °C, and 100 °C (Fig. S 1).

### *Data Analysis*

Bright-field TEM micrographs and SAED patterns were processed using Digital Micrograph (Gatan Microscopy Suite) and ImageJ (64). SED datasets were analysed using the open-source Python libraries HyperSpy (v.1.6.5 and 1.7.5) and pyXem (v.0.11.0 and 0.13.0) for multidimensional diffraction microscopy. Calibration was performed using standard $MoO_3$ and Au cross-grating samples as described previously (65). The rotation between the scan direction and diffraction pattern was corrected by measuring the angle between the long axis of $MoO_3$ crystals of known habit and corresponding diffraction spots. Pixel size calibration was achieved using an Au cross-grating with a 500 nm period, and residual elliptical distortions were removed by fitting diffraction rings from polycrystalline Au to ellipses that were then corrected to circles. For theophylline datasets, the direct beam was aligned to pixel precision and centred to subpixel accuracy using cross-correlation in pyXem. An affine transformation matrix was applied to correct the calibrated pattern rotation and elliptical distortion.

Diffraction patterns (originally $515 \times 515$ pixels) were cropped to $514 \times 514$ pixels and binned by a factor of two to produce $257 \times 257$ pixel patterns. Average electron diffraction patterns were generated by computing the mean intensity across all probe positions. Virtual dark-field (VDF) images were constructed in pyXem by integrating the intensity from selected regions of interest in the diffraction pattern. For visualisation, diffraction patterns were displayed as the square root of recorded intensity using ImageJ (64) to enhance contrast between high- and low-intensity features. Auto-indexing in CrystalMaker (v.10.7.3; 11.1.2; 11.5.1) was used to identify candidate zone axes. Simulated diffraction patterns were compared with calibrated, experimental diffraction patterns using structure files from the Cambridge Structural Database: CCDC 183780 and CCDC 2264675 for the monohydrate (49, 50) and CCDC 128707 for anhydrous form II (51).

Structural analysis was performed using CCDC Mercury (v.2024.3.1) (66), VESTA (v. 3.90.0a) (67), and CrystalMaker software (v.10.7.3; 11.1.2; 11.5.1). Water channel propagation and water space maps were modelled using the Hydrate Analyser tool, whilst predicted Bravais-Friedel-Donnay-Harker (BFDH) morphology was simulated in CCDC Mercury (v. 2024.3.1).

## Results

### *Morphology*

Our baseline starting sample comprised a theophylline hydrate prepared by crystallisation from a water-ethanol solvent mixture. Powder XRD confirmed the formation of theophylline monohydrate, with no residual anhydrous form II detected (Fig. 2). The experimental pattern matched the two reported monohydrate reference structures ($P2_1/n$ and $P2_1$), including characteristic reflections at 8.905° (011) and 14.787° (012) $2\theta$ (marked by asterisks), consistent with the monohydrate phase (Fig. 2a) (47). The pattern exhibits a minor reduction in lattice parameters relative to the reported structures, together with a reduced fit at peaks centred at 26-27° $2\theta$ and peak broadening, suggesting reduced crystallinity and disorder at hydrogen sites in either fitted structure, commonly attributed to sample preparation and preferred orientation (68).

Distinction between the closely related monoclinic $P2_1/n$ and $P2_1$ structures was not definitive from powder XRD alone, as both models produced near-identical Rietveld fits ($R_{wp}$ 5.135 vs 5.404% and GOF 3.495 vs 3.678, respectively; Table S 3). This is consistent with their nearly identical packing and the principal structural differences arising from hydrogen positions within the hydrogen-bonded water chains, which are poorly resolved by powder XRD (49). Additional reflections predicted for $P2_1$ (001 and 100; Table S 2, highlighted in orange) were negligible in intensity ($I/I_{max} < 0.01\%$), making them unlikely to be detected experimentally by powder XRD. Previous single-crystal studies assigned the monohydrate structure to $P2_1/n$ (50), with hydrogen disorder in the water chains associated with inversion symmetry, whereas later subgroup refinement proposed $P2_1$ as a plausible alternative based on theoretical calculations of the lowest energy structure (49). Our refinement marginally favoured $P2_1/n$; however, both structural models were considered in our subsequent electron diffraction analysis.

BF-TEM and SAED revealed thin, faceted, needle-like theophylline monohydrate crystals suitable for electron microscopy (Fig. 2b). This habit is expected for crystals obtained by aqueous recrystallisation and reflects the strong intrinsic anisotropy of the monohydrate lattice (46). The crystal structure consists of $\pi$-stacked dimer layers of molecules and aligned along the *a*-axis, with one-dimensional zigzag water channels (Fig. 1c). These structural motifs promote preferential growth along the *a*-axis, resulting in needle-like morphologies across a range of solvent systems (46, 54, 69). On the basis of the one-dimensional water-channel geometry, several reports have indicated that dehydration proceeds preferentially along the *a*-axis through diffusion and escape of water at the channel ends (19, 45, 70). No comparable channels are evident in other crystallographic directions.

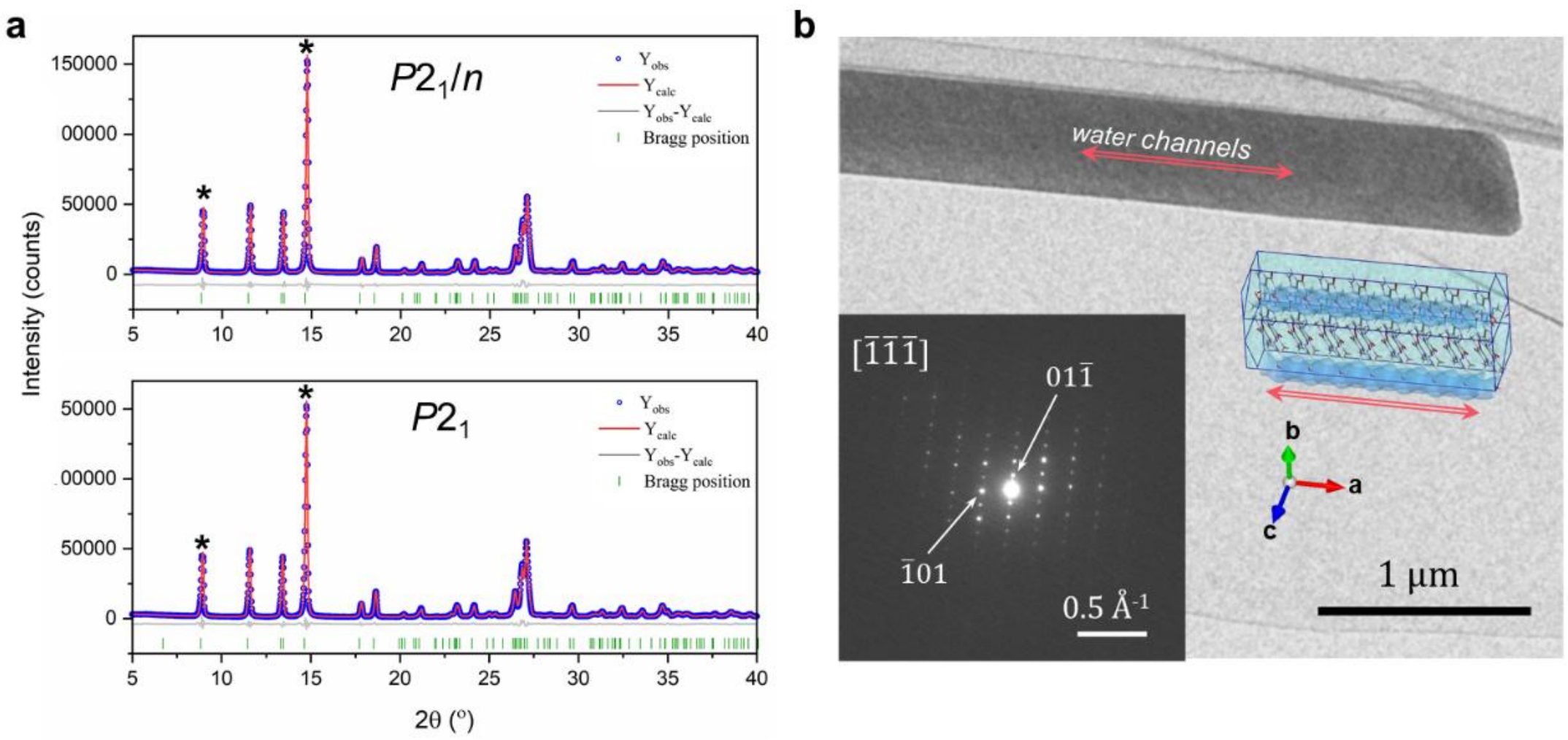

*Fig. 2. Theophylline monohydrate recrystallised from an ethanol-water mixture.* ***a*** *Rietveld refinement showed good agreement between the experimental and calculated patterns of the closely related $P2_1/n$ and $P2_1$ structures, confirming the monohydrate phase, although powder XRD alone cannot reliably distinguish between phases. The observed pattern ($Y_{obs}$, blue circles), calculated profile ($Y_{calc}$, red solid line), and difference curve ($Y_{obs}$ - $Y_{calc}$, grey solid line) are shown along with Bragg reflection positions, which are indicated by green vertical ticks. Diffraction peaks unique to the monohydrate $P2_1/n$ and $P2_1$ phases and not any anhydrous phases are marked with asterisks.* ***b*** *BF-TEM image showing needle-shaped crystals with SAED pattern and simulated BFDH morphology inset. Water molecules (highlighted blue regions in the simulated morphology) form zigzag channels along the a-axis, the fastest crystal growth direction.*

### *Experimental condition 1: In situ vacuum dehydration in the TEM column over time ($\Delta t$)*

Under condition 1, dehydration is induced solely by exposure to the high-vacuum environment of the TEM, without the application of additional heating. Under these conditions, vacuum-induced dehydration of theophylline monohydrate was investigated over time ($\Delta$t) by keeping samples in the TEM column, where the high-vacuum environment (~ $10^{-5}$ Pa) naturally promotes water loss. Under relatively mild dehydration conditions, such as vacuum drying at ambient temperature, the formation of the metastable form III prior to conversion to the stable anhydrous form II has been reported (45, 46, 48). Elevated temperature accelerates the transformation of form III to form II, while increased humidity has been shown to enhance the rate of solid-state conversion between these forms, resulting in a shorter lifetime of intermediate phases (46, 48). In the present *in situ* TEM experiments, dehydration occurs under high-vacuum within the microscope column, where humidity is very low. The distinct experimental environment here may also affect these dehydration pathways and influence the stability and detectability of intermediate phases.

To assess structural changes induced by prolonged vacuum exposure under condition 1, two SED datasets were acquired from the same regions of interest on selected theophylline crystals. The first dataset, collected after <30 min in vacuum, serves as a reference for the as-prepared state and provides the initial benchmark for subsequent analysis. A second dataset was acquired after ~17 h of continuous vacuum exposure to probe the effects of extended dehydration. Prior scoping studies indicated no significant changes after 1-2 h in vacuum; therefore, extended exposure was selected to maximise dehydration whilst maintaining efficient instrument use (i.e. the sample was kept in microscope vacuum overnight). The combined fluence from both scans (total electron fluence <16 $e^-$ $Å^{-2}$ i.e., <8 $e^-$ $Å^{-2}$ per scan) remained well below the critical electron fluence at room temperature and 300 keV for both anhydrous and monohydrate theophylline ($36 \pm 8$ $e^-$ $Å^{-2}$ and $29 \pm 9$ $e^-$ $Å^{-2}$, respectively (63, 71)).

Fig. 3a shows an ADF-STEM image of a representative crystal acquired after <30 min in vacuum, with representative diffraction patterns from the inner (purple squares) and outer (yellow squares) regions of interest (ROIs) used for subsequent analysis. Indexed diffraction patterns from both ROIs confirm the monohydrate phase oriented along $[111]$. Corresponding SED datasets acquired from the same regions after 17 h of continuous vacuum exposure show no additional diffraction spots or changes in reflection positions (Fig. 3b), while the overall needle morphology and diffraction contrast remain unchanged. This indicates no evidence of conversion to anhydrous form II. However, despite the absence of bulk structural changes, localised changes in crystal surface morphology are observed after overnight vacuum exposure (Fig. 3b, marked by blue arrows). Surface roughening is observed on one side of the needle, while the opposing surface remains smooth and well defined. More examples showing this behaviour are presented in Fig. S 2. The affected region shows directional material removal from the side of the needle along the length of the crystal. In contrast, Ren, S. *et al.* (72) report more general cracking and loss of optical transparency during freeze-drying of theophylline monohydrate, with formation of metastable form III crystals occurring across multiple surfaces. Theophylline form III could not be conclusively identified in our study, because its simulated diffraction pattern only differs from that of the monohydrate primarily in relative spot intensities, which are not reliably accessible here due to, e.g. crystal bending and/or dynamical electron scattering.

The formation of these surface features is consistent with localised mass loss from the crystal surface and may be described as an etching-like process induced by prolonged vacuum exposure. However, the surface features are not aligned with the water channel direction $[100]$ and instead propagate obliquely across the crystal surface (closer in alignment to the molecular $\pi$-$\pi$ stacking). This suggests that dehydration is not governed solely by simple axial out-diffusion of channel water and indicates a more complex interaction between surface processes and channel dehydration. At the needle ends, localised surface roughening is also observed around water channel openings; however, these features are less developed than those observed along the sides of the needle (Fig. S 2). Specifically, the roughened regions at the needle ends are both shallower in depth and less extensive in lateral coverage compared with the features at the needle sides (Fig. S 2, blue arrows for needle sides, red arrows for needle ends). This evidence of reduced material removal at the needle ends suggests slower water loss from the channels ends. The zigzag geometry along the channels and the smaller effective cross-sectional area at the needle ends may limit water transport and explain the observed persistence of the monohydrate phase under TEM vacuum conditions.

To investigate the crystallographic origin of the asymmetric surface roughening at the sides of the needle crystals, we modelled the BFDH morphology in the same orientation as the $[111]$ indexed diffraction pattern, i.e. viewed along the $[111]$ direction in the real-space unit cell, using both reported monohydrate crystal structures ($P2_1/n$ and $P2_1$) (49, 50). However, only the non-centrosymmetric $P2_1$ crystal structure provided a plausible structural explanation for the observed morphological changes (Fig. 3c). Experimentally, surface roughening was observed along only one of the two long needle sides (the bottom surface; Fig. 3b). In the corresponding BFDH morphology (Fig. 3c, Fig. S 3), one terminating face intersects the side of the hydrogen-bonded water-channel network, whereas the opposing face does not. Specifically, the $\{0\bar{1}1\}_A$ plane termination provides surface access to the water channel sides, while the opposing $\{0\bar{1}1\}_B$ plane termination does not, where A and B denote the polar and non-polar opposing surface terminations, respectively. Therefore, the roughened bottom particle surface can be assigned to the $\{0\bar{1}1\}_A$ terminating plane, while the comparatively smooth opposing side to the $\{0\bar{1}1\}_B$ termination. This asymmetric surface behaviour is only structurally rationalised by the non-centrosymmetric $P2_1$ crystal structure and provides a crystallographic basis for the selective surface roughening observed experimentally (Fig. 3d). Accordingly, the monohydrate phase is indexed using the $P2_1$ unit cell hereafter in the manuscript.

Thus, dehydration is apparently not governed solely by axial diffusion along the *a*-axis, but instead involves a coupled process of changes initiated on surfaces with exposed water channel sides $\{0\bar{1}1\}_A$, followed by progressive loss of channel water. These observations are consistent with previously reported TGA data and two-step dehydration models for theophylline monohydrate (19, 47). We note that the crystal shows a visible fault separating the thicker lower region and the thinner upper region (Fig. S 4). However, diffraction patterns from both areas confirm the orientation continuity of the needle on either side of the fault as discussed above (see explanation in Fig. S 4; the central fault is a crystal that is rotated around the *a*-axis to ~$[010]$, but the regions on either side of this fault lie along $[111]$).

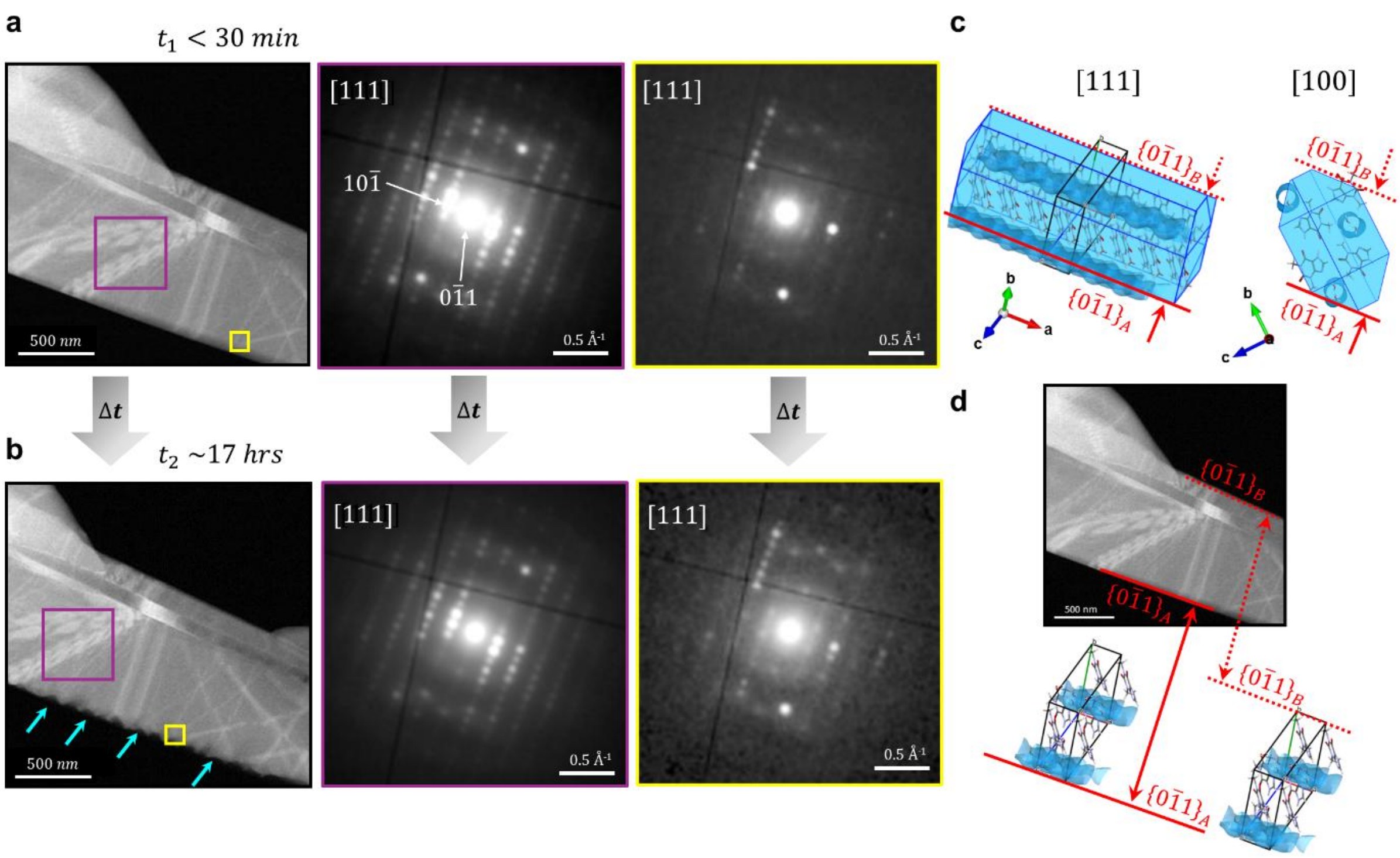

*Fig. 3. Experimental condition 1: In situ vacuum dehydration.* ***a*** *ADF-STEM image of a representative particle acquired after <30 min in vacuum, with corresponding SED patterns from the inner (purple square) and outer (yellow square) regions, both indexed to the monohydrate phase along the $[111]$ zone axis.* ***b*** *Corresponding SED datasets acquired from the same regions in* ***a*** *after 17 h of continuous vacuum exposure. No additional reflections or diffraction changes were detected, confirming preservation of the monohydrate phase. Characteristic surface roughening is indicated by blue arrows.* ***c*** *Corresponding BFDH morphology of theophylline monohydrate (using the non-centrosymmetric $P2_1$ crystal structure, (49)) oriented to the indexed $[111]$ zone axis, with an additional end-view projection corresponding to $[100]$. BFDH analysis indicates that the bottom crystal surface of the needle exposes terminating water channel sides and can be indexed to the $\{0\bar{1}1\}_A$, whereas the opposing top surface does not and corresponds to the $\{0\bar{1}1\}_B$ plane.* ***d*** *This structural anisotropy provides an explanation for the preferential roughening of the bottom surface, where water channel sides terminate at the $\{0\bar{1}1\}_A$.*

### *Experimental condition 2 - In situ vacuum heating dehydration ($\Delta T$)*

*In situ* heating experiments were performed under high-vacuum (condition 2, $\Delta$T) to activate thermally driven dehydration pathways (43, 47, 48) and to assess how elevated temperature influences dehydration kinetics, morphology, and spatial heterogeneity relative to vacuum exposure alone (condition 1, $\Delta$t).

Preliminary *in situ* BF-TEM imaging and SAED acquired during stepwise, nominally isothermal heating show that, at lower temperatures, the morphological changes closely resemble the changes observed under condition 1; surface roughening is likewise localised to specific crystal sides, followed by progressive mass loss consistent with surface-controlled dehydration (Fig. S 5). At higher temperatures (60 - 110 °C), additional morphological features emerge, including more extensive material removal in trenches under the etch pits that ultimately lead to pillar formation. In turn, SAED indicates phase transformation towards anhydrous form II with residual monohydrate reflections remaining detectable in the exposed pillars (Fig. S 6). These findings are consistent with reports of incomplete or spatially heterogeneous dehydration (43, 72). At the highest temperature range (100 - 110 °C), particle displacement or loss from the field of view is observed, likely enhanced by charging under electron-beam exposure (Fig. S 6). Because SAED averages diffraction over comparatively large regions, it does not allow direct correlation between local morphological changes and phase identity, motivating the use of spatially resolved SED to probe heterogeneous dehydration pathways during *in situ* heating.

For *in situ* heating experiments, an initial room temperature SED dataset was acquired on a selected theophylline monohydrate crystal, providing a reference for the as-prepared state and serving as the benchmark for subsequent measurements. SED scans were also acquired at 65 °C, 85 °C, and 100 °C on ROIs located close to, but not identical with the initial region (Fig. 4a). This acquisition setup was chosen to avoid cumulative beam damage and to ensure unexposed areas remained intact for comparison (73). The total electron fluence of the whole area was kept below 32 $e^{-}$ Å$^{-2}$ (accounting for <8 $e^{-}$ Å$^{-2}$ per scan). Despite the low-dose acquisition conditions, many particles fragmented or detached from the support film during the third and fourth scans (at 85 °C and 100 °C). The observation of similar behaviour on both $SiN_X$ window supports (used here for SED) and holey carbon films (initial *in situ* BF-TEM; Fig. S 6) suggests that the instability is predominantly temperature-driven.

Between room temperature and 65 °C, the particles exhibit surface roughening with a preferential, directional etching of specific crystal facets (purple arrows, Fig. 4a). Over this temperature range, diffraction patterns remain unchanged and can be indexed to $[\bar{5}1\bar{2}]$ of $P2_1$ crystal structure of theophylline monohydrate (Fig. S 7, S 8, S 9a and Table S 4) suggesting the etching of $\{0\bar{2}1\}$ crystal facets that lie close to the $\{0\bar{1}1\}_A$ polar planes (Fig. S 9b). The observed behaviour closely resembles that seen during *in situ* vacuum dehydration (condition 1; Fig. 3), suggesting a mass-loss-driven surface dehydration process common to condition 1 and condition 2. At 85 °C, directional etching becomes increasingly localised, leading to well-defined trench-like features, indicating that heating under vacuum accelerates mass loss. Diffraction patterns show localised blurring and weakening intensity of some reflections (orange arrows, Fig. 4a), consistent with increasing structural heterogeneity that might even suggest overlapping contributions from more than one phase. Upon heating to 100 °C, extensive and highly anisotropic material removal appears in the form of trench formation leading to the exposure of a continuous pillar framework with pillars oriented normal to the original, unetched facet of the crystal. At this stage, pronounced changes in the diffraction pattern are observed: reflections characteristic of the monohydrate phase are no longer present (blue arrows), while new, systematically shifted reflections appear (red arrows) (Fig. S 7). These diffraction features can be indexed to anhydrous theophylline form II oriented along the $[\bar{1}03]$ zone axis (Fig. S 10 and Table S 4). Comparisons of overlaid diffraction patterns from the monohydrate and anhydrous form II highlight subtle but systematic changes in reciprocal-space geometry, confirming the phase transformation (see also Fig. S 11i).

Fig. 4b-c show overlays of the same crystal before and after heating to 100 °C. Morphological changes during *in situ* heating and dehydration increased charging susceptibility, resulting in a small

rigid body displacement of the particle, including a ~3° rotation between the room temperature and 100 °C datasets (Fig. 4c). This rotation is already present at the start of the 100 °C acquisition and is therefore attributed to thermal repositioning of the sample. The displacement is consistent across both real-space images and diffraction patterns, confirming that it reflects rigid-body motion of the crystal on the grid.

When the pre- and post-heating diffraction patterns are aligned by rotation alone, multiple sets of reflections exhibit clear similarity between the monohydrate and anhydrous form II (Fig. 4d). This indicates that several crystallographic planes retain similar orientations throughout the dehydration. Although some reflections related to the monohydrate phase disappear upon transformation, a subset of reflections remains aligned along common reciprocal lattice directions, demonstrating that planes with similar orientations persist in both structures. These corresponding reflections occur at similar angular positions but are systematically shifted to higher scattering angles in anhydrous form II, consistent with its reduced *d*-spacings (see zoomed overlays in Fig. S 11i and detailed analysis in Fig. S 12).

The preservation of multiple plane families with common orientations before and after dehydration defines a specific orientation relationship between the monohydrate and anhydrous form II. This behaviour provides clear evidence that dehydration proceeds via a topotactic solid-state transformation, in which partial lattice continuity is maintained while molecular rearrangement is required to accommodate the phase change, including the inherent mass loss accompanying dehydration. The following section examines the crystallographic constraints and plausible transformation pathways that are consistent with this behaviour.

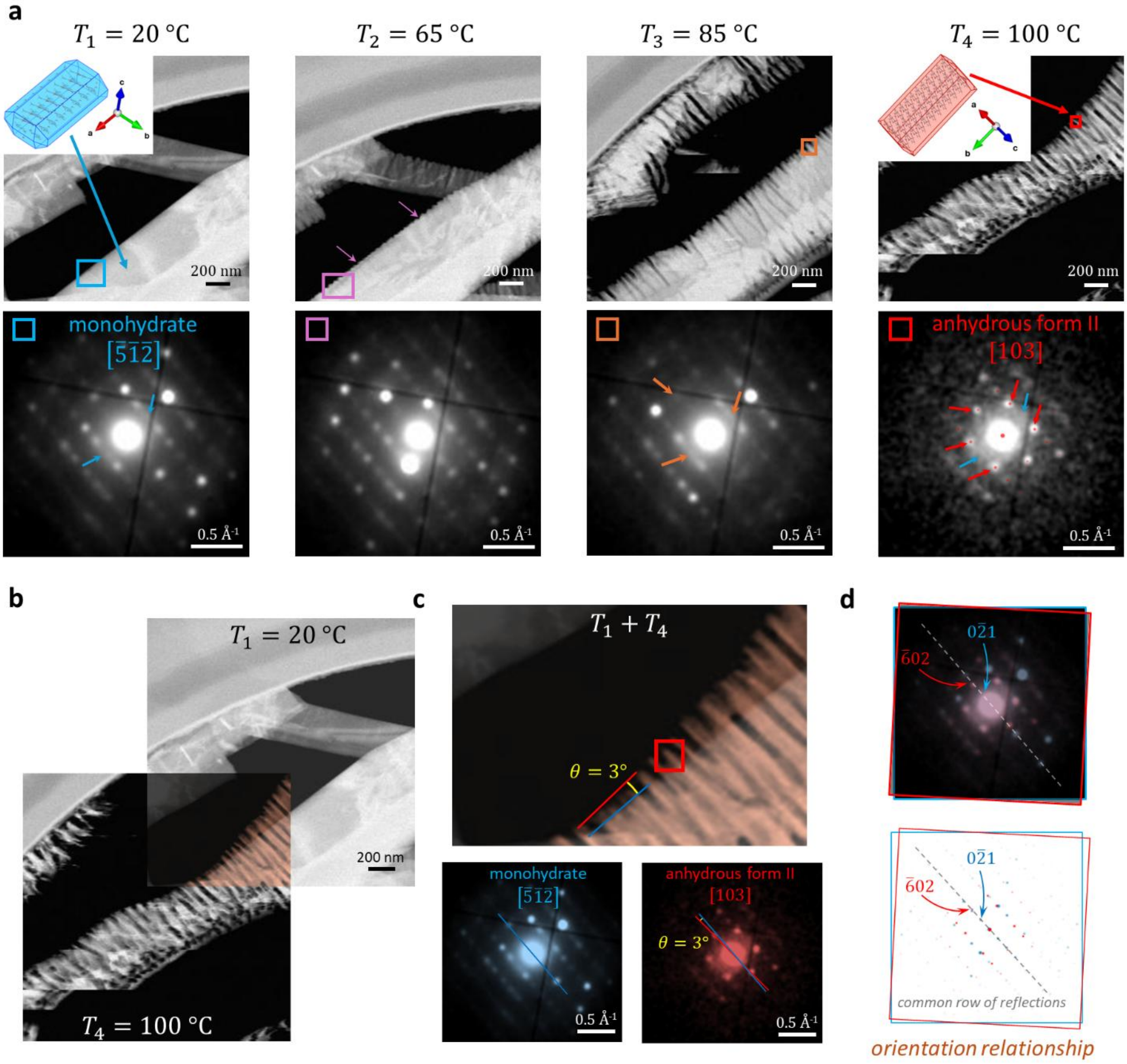

*Fig. 4. Experimental condition 2: in situ vacuum heating from 20 °C to 100 °C.* ***a*** *Progressive surface roughening is observed close to the* $\{0\bar{1}1\}_A$ *polar planes (Fig. S 9b), followed by the formation of pillar-like monohydrate domains, which convert into a skeletal framework of anhydrous form II at 100 °C. Selected regions corresponding to the presented diffraction patterns are marked by coloured squares: 20 °C (blue), 65 °C (pink), 85 °C (orange), and 100 °C (red). Corresponding arrows indicate diffraction features: blue arrows mark reflections present at 20 °C that disappear at 100 °C; orange arrows indicate reflections that broaden at 85 °C; red arrows indicate reflections shifted at 100 °C.* ***b*** *Overlay of pre- and post-heated regions analysed by SED.* ***c*** *Magnified view of the selected area showing a 3° relative tilt in real space, corroborated by corresponding electron diffraction patterns. Blue lines indicate the crystal side and common row of reflections at 20 °C (monohydrate), whereas red lines indicate the crystal side and common row of reflections at 100 °C (anhydrous form II).* ***d*** *Comparison of experimental and simulated diffraction patterns confirming the rotation between the monohydrate and anhydrous form II and highlighting the common row of reflections (blue for monohydrate and red for anhydrous form II) between the two phases.*

## *Structural modelling*

To examine the molecular scale rearrangements required to accommodate topotactic dehydration as inferred under condition 2 (Fig. 4), we modelled the crystal structures of theophylline monohydrate and anhydrous form II and directly compared these with the experimentally observed diffraction relationships. This analysis is undertaken to assess whether a physically plausible molecular reorganisation can account for the observed crystallographic relationship, to explore whether formation of form II occurs simultaneously with mass loss or may be partially decoupled from it, and to identify the specific molecular rearrangements required to accommodate the preservation of crystal orientation. Such considerations are essential for interpreting the transformation and for identifying structural features that may enable control over the dehydration process.

Both model structures were plotted such that the planes described by the common row of diffraction reflections are aligned (i.e. reflections $0\bar{2}1$ for monohydrate and $\bar{6}02$ for anhydrate form II, labelled in Fig. 4d). For the purposes of presentation in Fig. 5a, the corresponding real-space molecular models were displayed by rotating both until the anhydrous form II *b*-axis is horizontal on the page. This choice of representation facilitates direct visual comparison between the preserved orientation relationship and changes in molecular packing.

Direct identification of shared structural motifs by superposition of the two structures is hindered by the complexity of overlapping molecular projections. To address this, we simplify the representation and consider also a second view with both structures rotated to visualise the conjugated ring-system faces (Fig. 5b). In simplifying, we focus on a reduced description of the structural change, constrained by the experimentally observed orientation relationship between the monohydrate and anhydrous form II. The transformation pathway is modelled under the assumption of minimal molecular rearrangement (minimum mass transfer or molecular diffusion), corresponding to the shortest plausible geometric pathways between the initial and final configurations. These conditions minimise molecular displacement and transient volume. The theophylline molecule comprises a fused six-membered and five-membered (imidazole) ring system sharing a common C=C bond, forming an extended conjugated, approximately planar $\pi$-bonded framework. This geometry defines a unique molecular plane and a long molecular axis passing through the carbon of the imidazole ring. Two carbonyl groups and two methylamine moieties on the six-membered ring rigidly break mirror symmetry, thereby defining a second, orthogonal in-plane axis perpendicular to the long axis.

These key in-plane, molecular orientation descriptors can be used to generate simplified structural descriptions for each phase (Fig. 5c (i)): (1) first, the molecular long axis is marked, indicated by the direction passing through the apical imidazole carbon (represented by a line and dot), and (2) next, the orientation of functional groups within the molecular plane is marked, particularly the direction of the carbonyl group closest to the imidazole ring (represented by an arrowed line perpendicular to the long axis, with the arrow indicating the C=O bond direction) (Fig. 5c (i)).

Both the monohydrate and anhydrous form II structures consist of columns of theophylline molecules arranged through $\pi$-$\pi$ stacking. In these stacks, the molecular plane is inclined with respect to the stacking axis, giving rise to a well-defined out-of-plane molecular tilt that is determined by crystal packing rather than the intrinsic molecular geometry. As a result, distinct sequences are apparent where neighbouring stacks may exhibit either the same molecular tilt orientation relative to the stacking direction or opposing tilts. To quantify this out-of-plane stacking behaviour, we define an acute angle, $\theta$, between the stacking direction (defined by the linear arrangement of molecules within a column) and a crystallographic reference axis, taken here for ease of viewing as the *c*-axis of the monohydrate and *a*-axis of the anhydrous form II. For convention, $\theta > 0$ and $\theta < 0$ denote the two opposite orientations of the stacks, labelled as ‘up’ (u) and ‘down’ (d), respectively. This definition is illustrated in Fig. 5c (ii) using side-view, real-space structure models displayed in a plane containing both the stacking directions (yellow for ‘down’ and purple for ‘up’) and the reference axis (green line), which is oriented horizontally for clarity. A key structural distinction between the two phases emerges from this description. The monohydrate exhibits a uniform stacking arrangement in which all molecules adopt the same out-of-plane orientation (either all ‘up’ or all ‘down’). In contrast, anhydrous form II displays an alternating stacking sequence, such as ‘up, down, down, up’ or its inverse, reflecting a systematic alternation of molecular tilt between adjacent stacks (Fig. 5c (ii)).

All possible simplified geometric descriptor configurations of a theophylline molecule in the two phases are summarised in Fig. 5d. For each phase, four configurations are considered, arising from combinations of the two out-of-plane molecular tilt orientations (‘up’ and ‘down’) and the two possible in-plane orientations of the carbonyl group closest to the imidazole ring relative to the molecular long axis. These simplified descriptors can then be mapped onto each unit cell by replacing the full molecular representations with the corresponding geometric features. To enable this comparison, the unit cells of the two oriented phases shown in Fig. 5a were both rotated anticlockwise, within the same viewing plane, by 70° about a vertical axis to provide a view along the molecular stacking direction (Fig. 5b). In this orientation, application of the simplified geometric descriptors allows direct, site-by-site comparison of the molecular arrangements in the two phases.

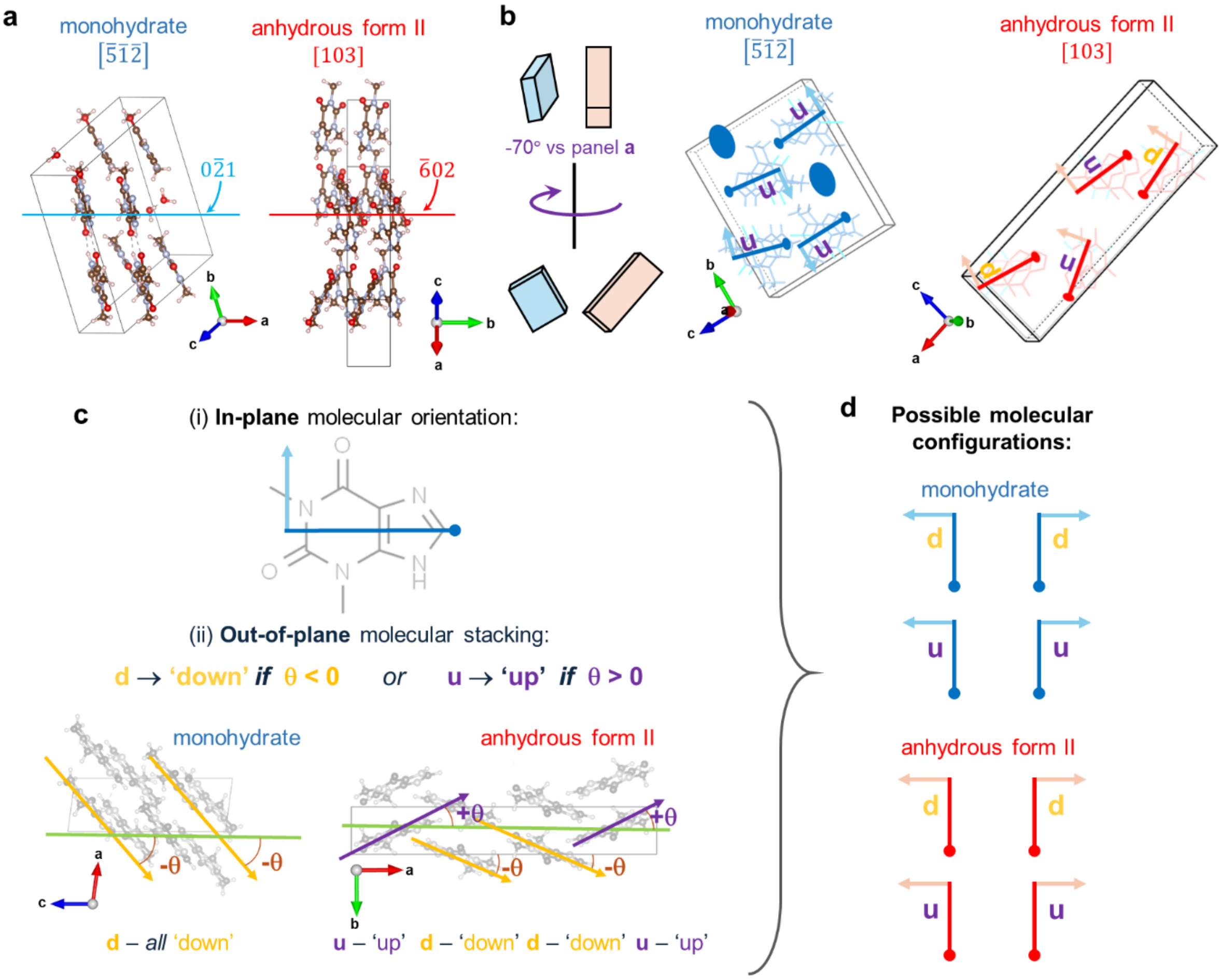

*Fig. 5. Structural modelling to interpret the observed phase transformation. **a** Relative viewing orientation of the unit cells of the theophylline monohydrate and anhydrous form II determined by the in situ heating experiment. Simulated crystal structures of the monohydrate and anhydrous form II were aligned with the corresponding experimental diffraction patterns and displayed by rotating both until the b-axis of anhydrous form II is horizontal to facilitate visual comparison. **b** The same unit cells rotated horizontally relative to panel (a) to provide a view along the molecular chains, with the simplified structural features applied to enable direct comparison between the two structures. **c** Key geometric descriptors defining the (i) in-plane molecular orientation and the (ii) out-of-plane stacking used for the simplified structural representation. **d** All possible simplified geometric descriptor configurations for the theophylline phases, representing all molecular orientation and stacking combinations (shown in **c** (i) and **c** (ii) i.e. that enable direct molecular comparison between the monohydrate and anhydrous form II.*

We systematically evaluated all experimentally consistent zone-axis orientations identified from the *in situ* heating experiment to identify viable crystallographic relationships underlying the phase transformation. These specific zone axes were selected based on the criterion that they reproduce the experimental diffraction pattern geometry and spacings, including the small but systematic shifts of diffraction spots observed between the monohydrate and anhydrous phases (Fig. S 11i,ii). The analysis considered two criteria. First, the allowed permutations of the monoclinic zone axis $\langle 512 \rangle$ and $\langle 521 \rangle$ that match the experimental diffraction pattern of monohydrate give 4 zone-axis orientations plus the 4 allowed permutations of the orthorhombic zone axis $\langle 103 \rangle$ that match the experimental diffraction pattern of anhydrous form II to give a total of 16 possible zone-axis combinations between the two phases. Second, relative rotational alignment between the monoclinic and orthorhombic unit cells was considered, allowing for both 0° and 180° rotations of the Bragg reflections, resulting in 32 candidate configurations overall (Table S 4).

Each candidate configuration was evaluated following the procedure illustrated in Fig. S 11. First, candidate zone-axis pairs were screened by comparing simulated diffraction patterns, retaining only 16 configurations consistent with the small outward shift of diffraction spots observed for anhydrous form II relative to the monohydrate in the experimental diffraction patterns (see the zoomed overlay in Fig. S 11i). Second, a simplified representation of each unit cell was applied, replacing molecular structures with the geometric descriptor features (Fig. S 11ii). Third, molecular stacks in anhydrous form II oriented in the same direction as those in the monohydrate phase were identified (Fig. S 11iii). In the monohydrate, all molecules within a given stack have the same orientation (either all 'up' or all 'down'); therefore, only equivalently oriented molecules were considered in anhydrous form II. Within these stacks, preservation of molecular asymmetry across the phases was imposed as a constraint by requiring that the relative orientation of functional groups within the molecular plane be maintained and not flipped. Common molecular orientation pairs within each candidate zone axis configuration were identified with a common motif, defined by the identical in-plane orientation of the C=O group closest to the imidazole ring in both phases, using an extended unit cell view (Fig. S 11iv). Finally, the in-plane rotation required to achieve full alignment of these molecular pairs between the two phases was measured in the plane of the visualisation of the registered simplified model. Configurations requiring rotations exceeding 90° were excluded, leaving only pairs involving the smallest rotational adjustments (Fig. S 11v).

After evaluating the original 32 potential configurations (Table S 4 and Fig. S 9, 10), we narrowed the possibilities to four viable candidates that shared a common molecular motif, identifying the same equivalent planes across both phases i.e. $(001)$ for monohydrate and $(100)$ for anhydrous form II (Fig. S 13). Of the four identified options, the configuration of $[\bar{5}\bar{1}\bar{2}]$ for monohydrate and $[103]$ for anhydrous form II required only a small rotation of molecules located across a common intermolecular plane of each crystal; also, this shared motif lay directly on the $(001)$ monohydrate surface while remaining confined within a single anhydrous form II unit cell, making it the most plausible pathway (Fig. 6a). The two model structures were aligned through the common planes and then rotated into a side view to clearly illustrate the common molecular motif across both phases. This visualisation shows that the proposed mechanism involves (i) nucleation of anhydrous form II at the monohydrate surface followed by (ii) topotactic crystal growth (Fig. 6b). The structures were then rotated back clockwise by 70° within the viewing plane relative to panel **a** to match the zone axes observed experimentally in the diffraction patterns (Fig. 6c), revealing that these identified common planes are nearly parallel and establishing a clear and consistent orientation relationship between the monohydrate and anhydrous form II. The transformation from monohydrate $[\bar{5}\bar{1}\bar{2}]$ to anhydrous form II $[103]$ is confirmed by diffraction pattern simulations (Fig. S 12).

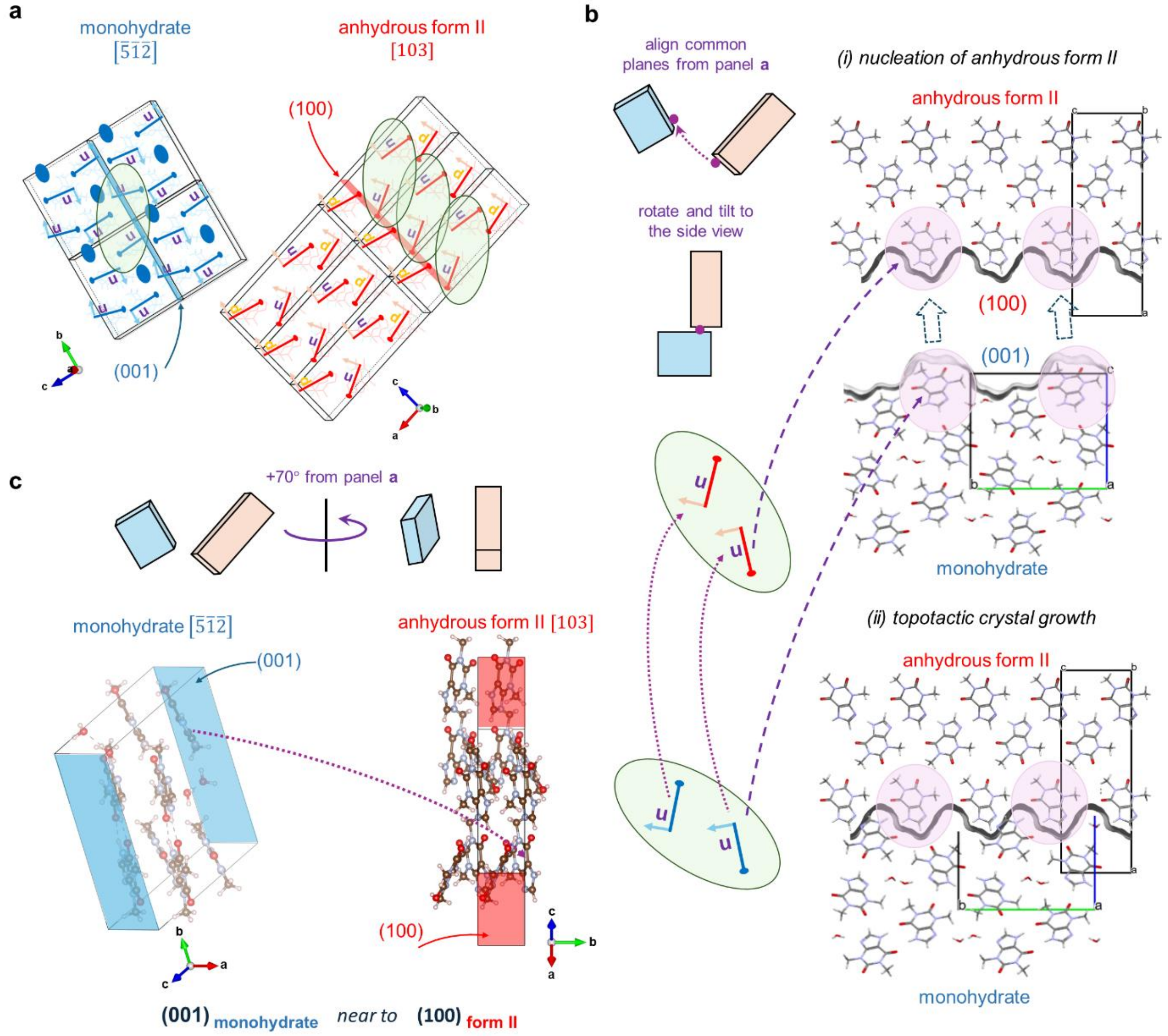

*Fig. 6. Orientation relationship between theophylline monohydrate and anhydrous form II.* ***a*** *Common structural motif located on the* $(001)$ *plane of the monohydrate and the* $(100)$ *plane of anhydrous form II.* ***b*** *Side view to clearly illustrate the common molecular motif across both phases.* ***c*** *Rotation of the two structures to match the zone axis observed in the experimental diffraction pattern (with the co-linear diffracting planes horizontal; Fig. 5a), reveals that these planes are nearly parallel, confirming a clear and consistent orientation relationship between the two polymorphs. The rotation angle is estimated from the simplified structural model.*

# Discussion

The model derived from the *in situ* low-dose SED measurements is summarised schematically in Fig. 7. As illustrated in the first stage of the figure, both experimental conditions reported here indicate that dehydration of theophylline monohydrate initiates at specific crystal surfaces corresponding to terminations of the water channel sides (at or close to the $\{0\bar{1}1\}_A$ of the polar $P2_1$ monohydrate structure), as well as at the end faces of the needles ($\{100\}$ of the same structure), where axial out-diffusion from water channel ends can occur (Fig. 3, Fig. 4). This initial stage is characterised by surface-localised dehydration at the sides of the long axis of monohydrate needles accompanied by surface roughening of these sides, mass loss, and preferential etching, demonstrating that water removal does not proceed solely by simple axial diffusion through open channels at the needle ends. This surface roughening reflects the combined influence of channel geometry, local packing constraints, and differential surface accessibility, which together govern the earliest stages of dehydration and are inferred to limit the overall rate of water removal. Such behaviour is consistent with the narrow cross-section and zig-zag topology of the water channels reported for theophylline monohydrate (19).

In the subsequent stage shown in Fig. 7, within the experimental time window studied, transformation beyond partial dehydration requires elevated temperature, as anhydrous form II is not detected under vacuum alone (condition 1). Under additional heating (condition 2), continued mass loss leads to increasingly pronounced directional etching and the exposure of a pillar-like surface topology, particularly evident along etched sides of monohydrate needles (Fig. 4). These pillars correspond to regions of the monohydrate crystal, or possibly a short-lived metastable hydrate, from which channel water has been partially removed. This behaviour is consistent with previous reports indicating that dehydration proceeds more rapidly near crystal surfaces while the particle interior remains hydrated for a longer period (42, 47, 53). Throughout this etching-dominated regime, SED patterns remain sharp and indexable to the monohydrate phase. At higher temperatures (85 °C), selected reflections exhibit slight blurring (Fig. 4a), which may arise from static or dynamical disorder associated with anisotropic mass loss, local strain, or superposition of diffraction from structurally heterogeneous domains. Importantly, long-range crystallinity is preserved. This observation complements interpretations based on vibrational spectroscopy that suggest transient amorphous or highly disordered states can occur under certain thermal or humidity conditions (44, 72). Under the *in situ* vacuum heating conditions employed here, dehydration proceeds through crystalline monohydrate with increasingly defective domains prior to crystallographic rearrangement to anhydrate, consistent with the two-stage desolvation model proposed by Perrier, P. and S.R. Byrn (19), in which initial weakening of the hydrogen-bonded framework precedes reconstructive transformation rather than amorphisation.

We do not resolve diffraction intensities that would allow the metastable form III to be identified unambiguously. This outcome is consistent with structural analyses by Paiva, E.M. et al. (45) and the initial stages of molecular-dynamics simulations by Larsen, A.S. et al. (70), which showed that reciprocal-space differences between the monohydrate and the metastable form III, the so-called 'dehydrated hydrate', are subtle and dominated by changes in relative diffraction intensities rather than by the appearance or disappearance of unique reflections. In electron diffraction, reliable interpretation of intensity variations is complicated by incomplete integration of reciprocal lattice intensities under kinematical conditions, making them highly sensitive to small orientation changes (e.g. crystal bending or mosaicity), and further by dynamical scattering, where multiple scattering pathways contribute to the observed intensity of a given reflection. While dynamical effects do not permit arbitrary redistribution of intensity, the superposition of multiple scattering routes can substantially modify relative reflection intensities, particularly in low-symmetry, beam-sensitive organic crystals such as the theophylline phases investigated here. Under these conditions, intensity differences expected between the monohydrate and metastable form III cannot be distinguished with confidence. Taken together with the morphological changes illustrated in Fig. 7, these considerations support the interpretation of the etched, pillar-like regions observed here as partially dehydrated monohydrate domains.

At the final temperature (100 °C) of the *in situ* heating experiment, corresponding to the final transformation step summarised in Fig. 7, we unambiguously identify anhydrous form II on the exposed pillar framework (Fig. 4). Analysis of the molecular packing reveals a common structural motif located on the $(001)$ plane of the monohydrate and the $(100)$ plane of anhydrous form II (Fig. 6). These planes share similarity in molecular arrangement and establish a clear orientation relationship between the two phases. This relationship is consistent with our *in situ* dehydration SED data analysis, i.e. once dehydration progresses beyond the initial mass-loss-driven surface etching, anhydrous form II nucleates at the surface of pillar-like domains and grows with an orientation that reflects this crystallographic relationship. The transformation is reconstructive in nature, involving reorganisation of the hydrogen-bonding network and adjustment of intermolecular distances. However, the lattice orientation is preserved across the phase boundary (Fig. 6b), indicating a topotactically-guided nucleation process. While our diffraction data do not provide evidence for an intermediate amorphous phase of appreciable volume or lifetime, we cannot exclude the presence of very small-scale or highly transient disordered regions during dehydration, for example, associated with local molecular rearrangement at retreating surfaces or molecular rearrangement at the interface between the monohydrate and the growing anhydrous phase. Any such disordered material must, however, be confined to short timescales and limited spatial extents, as long-range orientational coherence is retained in the final anhydrous pillars. Similarly, material removed during trench formation could, in principle, undergo short-range reorganisation and re-attachment to the growing form II surfaces, although our data do not allow this to be directly assessed.

Orientation relationships of this type have previously been proposed for theophylline in the context of hydration, where epitaxial growth of the monohydrate on anhydrous form II was reported using a $P2_1$ structure (41, 62). The monohydrate structure has since been reassigned to space group $P2_1/n$ (42, 50) with later subgroup analysis again supporting $P2_1$ as a plausible alternative (49). These structures are consistent with our Rietveld refinement and highlighting the limitations of the original Sutor, D. (62) space group assignment. Our reassessment of the orientation relationship proposed by Rodríguez-Hornedo, N. et al. (41) using the recently updated $P2_1$ monohydrate structure (49) (selected here because the observed preferential etching of specific crystal faces, e.g. in Fig. 3, is consistent with a non-centrosymmetric structural description) reveals that the same common planes to those identified in this study are present (Fig. 6, Fig. S 14). These results indicate that the dehydration pathway proceeds through defined crystallographic relationships between the two phases, despite mass loss and morphological change.

The morphological changes associated with this transformation closely resemble those reported in numerous *ex situ* studies. Surface roughening, crack formation, and loss of mechanical integrity have been demonstrated for dehydrated theophylline under a range of environmental conditions (42, 47, 54). Our *in situ* observations reveal the structural mechanism responsible for these changes: as dehydration accelerates because of cracking at the etched pits initiated at lateral terminations of the water channel sides, the nucleation of form II occurs preferentially on pillar surfaces. This surface reconstruction via topotactic nucleation (15)) produces discrete anhydrous domains of high packing density on top of partially dehydrated monohydrate regions across a plane of common molecular orientation, ultimately resulting in the fragmented morphologies observed in previous studies.

Macroscopic kinetic analyses of theophylline dehydration have traditionally relied on fitting experimental data to idealised mathematical models, including Avrami-Erofeev nucleation-growth, phase-boundary contraction, and diffusion-controlled mechanisms (39, 44, 52). These approaches yield widely varying Avrami exponents ($n \approx 0.25$-3), a variability attributed to differences in particle size, heating rate, sample geometry, and microstructure (52, 74). Our nanoscale observations can help explain this inconsistency. The early or slow stages of transformation are dominated by surface processes, including anisotropic channel collapse and limited nucleation sites with constrained growth of the anhydrous form, consistent with the low Avrami exponents reported for slow or low temperature monohydrate dehydration (43, 52). Once surfaces exposed to the top layer of water channels have largely dried, further dehydration is controlled by removal of water from subsurface hydrated domains, consistent with the diffusion-limited behaviour identified by Touil, A. et al. (53).

This results in slow inward progression of etching that ultimately leads to cracking at the head of the etch pits, significantly accelerating the dehydration process, i.e. transitioning to the second stage of the transformation. Any diffusion models that capture these higher dehydration kinetics of this second stage may give higher Avrami exponents because the growth of the strongly localised, orientation-guided nucleation of anhydrous form II can extend in more dimensions, i.e. raising the Avrami exponent to higher *n* (47).

Taken together (Fig. S 15), dehydration of theophylline monohydrate drives surface roughening at or close to $\{0\bar{1}1\}_A$ polar planes by mass loss that leads to preferential, facet-selective etching and localised trench formation. This surface roughening progressively exposes dehydrated monohydrate pillar domains that retain the orientation of the parent lattice. These exposed pillars then dehydrate more rapidly, and anhydrous form II nucleates at their surfaces with a well-defined crystallographic orientation relationship between $(001)$ of the monohydrate and $(100)$ of anhydrous form II. This is a reconstructive topotactic nucleation and crystal growth mechanism, in which substantial rearrangement of molecular packing occurs while a common molecular orientation is preserved across the interfacial plane, providing direct structural validation of transformation pathways previously only inferred from bulk studies. Furthermore, this basic mechanism may also have implications for dehydration processes in other molecular channel hydrates and even ionic inorganic hydrates such as the dehydration of gypsum to basanite which is proposed to proceed via a topotactic transformation (75).

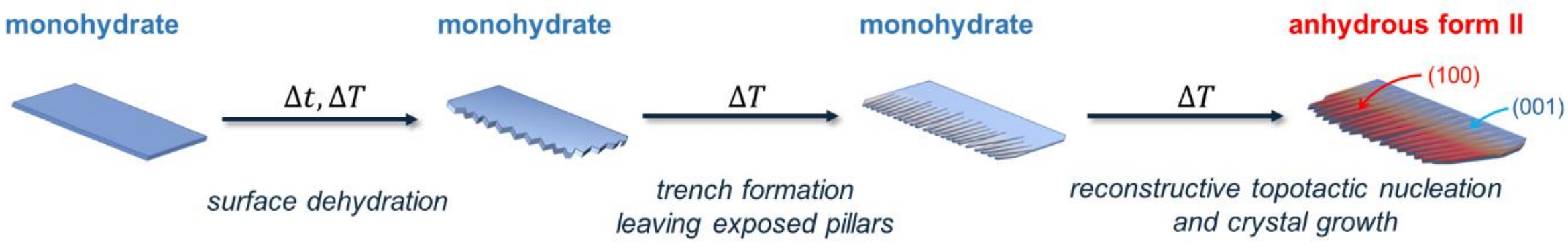

*Fig. 7. Data-driven model of phase transformation during dehydration of theophylline monohydrate. In situ low-dose SED shows that first monohydrate undergoes mass loss on polar facets that run parallel to water channels (such as monohydrate the $\{0\bar{1}1\}_A$ plane), these then etch into trenches that leave exposed pillars of monohydrate and secondly anhydrous form II nucleates on the surface of the exposed monohydrate pillars across crystallographic planes with common molecular orientation between the two structures; $(001)$ of the monohydrate and $(100)$ of anhydrous form II. This orientation relationship enables reconstructive topotactic nucleation and subsequent crystal growth of the anhydrous form II.*

## Conclusions

This study describes the microscopic sequence of structural changes associated with vacuum-driven dehydration of theophylline monohydrate and its solid-state transformation to anhydrous form II. Dehydration is thermally activated, occurring slowly at room temperature and accelerating strongly between 65 and 100 °C. Initial water loss is accompanied by anisotropic surface roughening, mass loss and etching at surfaces where water channel sides terminate (i.e. on facets close to the polar $\{0\bar{1}1\}_A$ monohydrate planes; supporting a non-centrosymmetric monohydrate structural model) and progresses through localised pitting and trench formation. These processes leave pillar-like monohydrate domains that preserve the orientation of the parent lattice. Further dehydration occurs preferentially within these exposed regions and leads to nucleation and growth of anhydrous form II on the monohydrate with a clear crystallographic orientation relationship, which arises from a shared molecular arrangement at the $(001)$ of monohydrate and $(100)$ of anhydrous form II. We identify the transformation as a reconstructive topotactic process in which molecular packing is reorganised while lattice orientation is retained.

These findings explain the fragmentation and loss of mechanical integrity reported previously during dehydration of theophylline crystals and resolve conflicting interpretations as to whether a direct solid-state transformation pathway is feasible. While the mechanism described here is specific to vacuum and thermally driven dehydration, it highlights how surface-controlled mass loss and crystallographic constraints control phase transformations in this and potentially other molecular hydrates. More generally, the ability to correlate nanoscale morphology with crystallographic changes provides a basis for designing strategies to inhibit or promote dehydration, for example, through surface modification or facet-selective treatments, and suggests new routes for controlling crystal orientation and microstructure in pharmaceutical solids and beyond.

## Acknowledgements

The authors acknowledge funding from AstraZeneca and the UK's Engineering and Physical Sciences Research Council (EPSRC) for an iCASE studentship (no. 2182593). The authors also acknowledge funding support from the EPSRC through the grant EP/X040992/1. We thank the Diamond Light Source, Rutherford Appleton Laboratory, UK, for access to the electron Physical Sciences Imaging Centre (ePSIC; MG28500, MG30057 and MG39047). We thank T. Slater and C. Allen for support at ePSIC. We also thank Prof. Kevin Roberts, Dr CaiYun Ma and Ms Gabriele Sumanskaite for helpful discussions on theophylline monohydrate structures.

## Supplementary Information

*Table S 1 Crystallographic information for theophylline monohydrate and theophylline anhydrous form II.*

| | **Theophylline monohydrate [a, b]** | | **Theophylline anhydrous form II [c]** |
|---|---|---|---|
| crystal system | monoclinic | monoclinic | orthorhombic |
| space group | $P2_1/n$ | $P2_1$ | $Pna2_1$ |
| $a$ (Å) | 4.468 (2) | 4.46150(10) | 24.612 (2) |
| $b$ (Å) | 15.355 (5) | 15.3156(3) | 3.8302 (4) |
| $c$ (Å) | 13.121 (5) | 13.0669(3) | 8.501 (5) |
| β (°) | 97.792 (7) | 98.529(4) | 90 |
| volume (Å$^3$) | 891.9 (6) | 885.111 | 801.38 (12) |
| density (g/cm$^3$) | 1.476 | 1.4873 | 1.493 |
| CCDC refcode | THEOPH01 | THEOPH08 | BAPLOT01 |
| CCDC deposition number | 183780 | 2264675 | 128707 |

[a] Sun, C., Zhou, D., Grant, D.J.W. and Young Jr, V.G. 2002. Theophylline monohydrate. Acta Crystallographica Section E. 58(4), pp.o368-o370.

[b] Konovalova, I.S., Shishkina, S.V., Wyshusek, M., Patzer, M. and Reiss, G.J. 2024. Supramolecular architecture of theophylline polymorphs, monohydrate and co-crystals with iodine: study from the energetic viewpoint. RSC Advances. 14(41), pp.29774-29788.

[c] Ebisuzaki, Y., Boyle, P.D. and Smith, J.A. 1997. Methylxanthines. I. Anhydrous Theophylline. Acta Crystallographica Section C. 53(6), pp.777-779.

*Table S 2 Simulation of the first 10 powder X-ray diffraction reflections (ordered by increasing 2θ) for theophylline monohydrate calculated for $P2_1/n$ and $P2_1$ space groups, listing corresponding Miller indices (hkl), d-spacings, and relative intensities to compare the closely related structures.*

| ***P2₁/n*** | | | | | ***P2₁*** | | | | |
|---|---|---|---|---|---|---|---|---|---|
| **h k l** | **d (Å)** | **2-Theta** | **Intensity** | **$I/I_{max}$** | **h k l** | **d (Å)** | **2-Theta** | **Intensity** | **$I/I_{max}$** |
| | | | | | **0 0 1** | 12.95338 | 6.818 | 2.36E-04 | **0.01%** |
| **0 1 1** | 9.92163 | 8.9051 | 1.10E+00 | 25.30% | **0 1 1** | 9.89033 | 8.9333 | 5.63E-01 | 24.60% |
| | | | | | **0 1 -1** | 9.89033 | 8.9333 | 5.63E-01 | 24.60% |
| **0 2 0** | 7.6775 | 11.5158 | 2.21E+00 | 50.80% | **0 2 0** | 7.6578 | 11.5456 | 2.23E+00 | 97.60% |
| **0 2 1** | 6.61071 | 13.3821 | 1.28E+00 | 29.50% | **0 2 1** | 6.59202 | 13.4202 | 6.43E-01 | 28.10% |
| | | | | | **0 2 -1** | 6.59202 | 13.4202 | 6.43E-01 | 28.10% |
| **0 0 2** | 6.49993 | 13.6112 | 7.64E-04 | 0.02% | **0 0 2** | 6.47669 | 13.6603 | 2.01E-03 | 0.10% |
| **0 1 2** | 5.98572 | 14.7868 | 2.59E+00 | 59.60% | **0 1 2** | 5.96524 | 14.8379 | 1.30E+00 | 56.90% |
| | | | | | **0 1 -2** | 5.96524 | 14.8379 | 1.30E+00 | 56.90% |
| **0 2 2** | 4.96081 | 17.8645 | 1.60E-01 | 3.70% | **0 2 2** | 4.94517 | 17.9215 | 8.98E-02 | 3.90% |
| | | | | | **0 2 -2** | 4.94517 | 17.9215 | 8.98E-02 | 3.90% |
| **0 3 1** | 4.76249 | 18.6149 | 5.95E-01 | 13.70% | **0 3 1** | 4.74963 | 18.6658 | 3.07E-01 | 13.40% |
| | | | | | **0 3 -1** | 4.74963 | 18.6658 | 3.07E-01 | 13.40% |
| | | | | | **1 0 0** | 4.42274 | 20.0592 | 1.03E-05 | **0.00%** |
| **1 0 -1** | 4.37537 | 20.2786 | 7.53E-01 | 17.30% | **1 0 -1** | 4.36472 | 20.3286 | 7.76E-01 | 34.00% |
| | | | | | **0 0 3** | 4.31779 | 20.552 | 5.89E-04 | 0.03% |
| **1 1 0** | 4.25351 | 20.866 | 8.29E-02 | 1.90% | **1 1 0** | 4.24912 | 20.8878 | 9.12E-02 | 4.00% |
| **-1 1 1** | 4.20788 | 21.0949 | 1.97E-01 | 4.50% | **1 1 -1** | 4.19759 | 21.1471 | 8.99E-02 | 3.90% |
| | | | | | **-1 1 1** | 4.19759 | 21.1471 | 8.99E-02 | 3.90% |

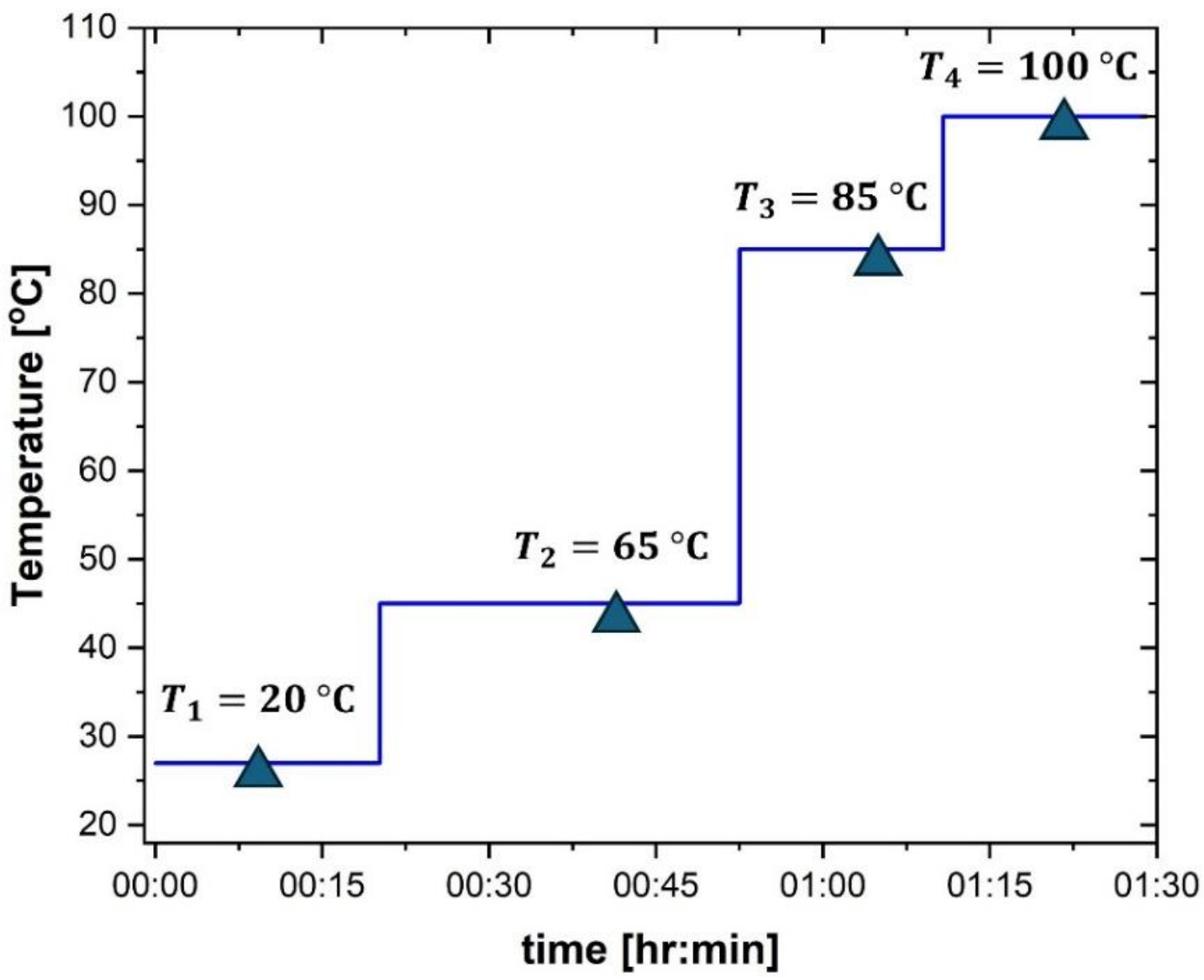

*Fig. S 1 Time-temperature profile of the in situ vacuum heating dehydration experiment performed using a ramped heating mode.*

*Table S 3 The quality of the Rietveld refinement was assessed using the conventional R-factors and by visual inspection of the agreement between the observed and calculated diffraction patterns. $R_{wp}$: weighted profile R-factor, GOF: goodness of fit.*

| | *Space group* | *a (Å)* | *b (Å)* | *c (Å)* | *β (°)* | *Vol (Å³)* | $R_{wp}$ *(%)* | GOF (%) |
|---|---|---|---|---|---|---|---|---|
| **Refined_a** | *$P2_1/n$* | 4.4976(3) | 15.4225(6) | 13.2913(4) | 98.535(4) | 911.74(8) | 5.135 | 3.495 |
| **THEOPH01[a]** | | 4.468(2) | 15.355(5) | 13.121(5) | 97.792(7) | 891.87 | - | - |
| **Refined_b** | *$P2_1$* | 4.4977(3) | 15.4223(6) | 13.2908(5) | 98.529(4) | 911.73(8) | 5.404 | 3.678 |
| **THEOPH08[b]** | | 4.46150(10) | 15.3156(3) | 13.0669(3) | 97.558(2) | 885.111 | - | - |

[a] Sun, C., Zhou, D., Grant, D.J.W. and Young Jr, V.G. 2002. Theophylline monohydrate. Acta Crystallographica Section E. 58(4), pp.o368-o370.

[b] Konovalova, I.S., Shishkina, S.V., Wyshusek, M., Patzer, M. and Reiss, G.J. 2024. Supramolecular architecture of theophylline polymorphs, monohydrate and co-crystals with iodine: study from the energetic viewpoint. RSC Advances. 14(41), pp.29774-29788.

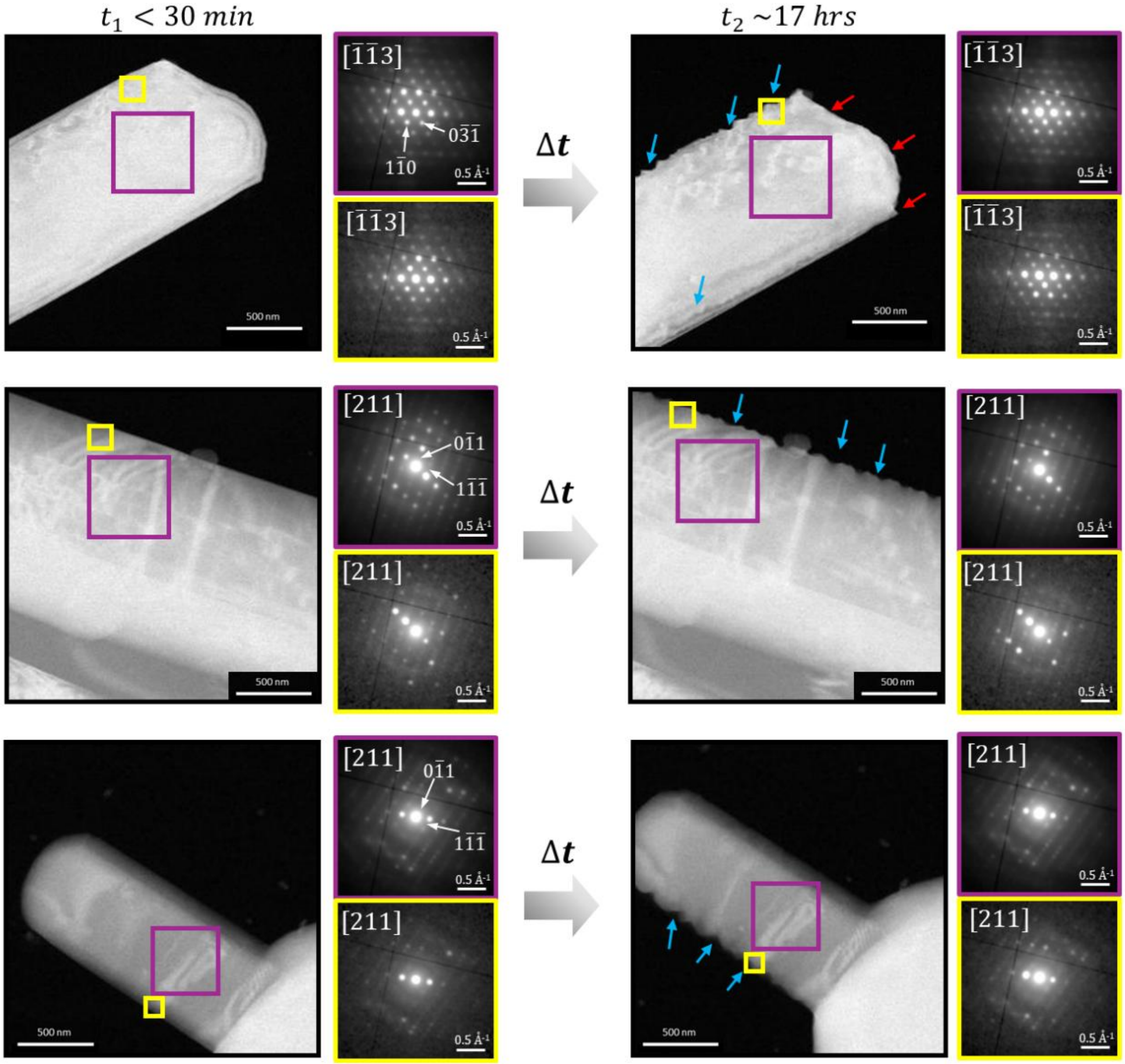

*Fig. S 2 Further examples of in situ vacuum dehydration under condition 1. Surface roughening is observed on specific crystal planes after ~17 h under high vacuum. Blue arrows highlight characteristic surface roughening at the crystal sides, and red arrows indicate roughening at the crystal ends.*

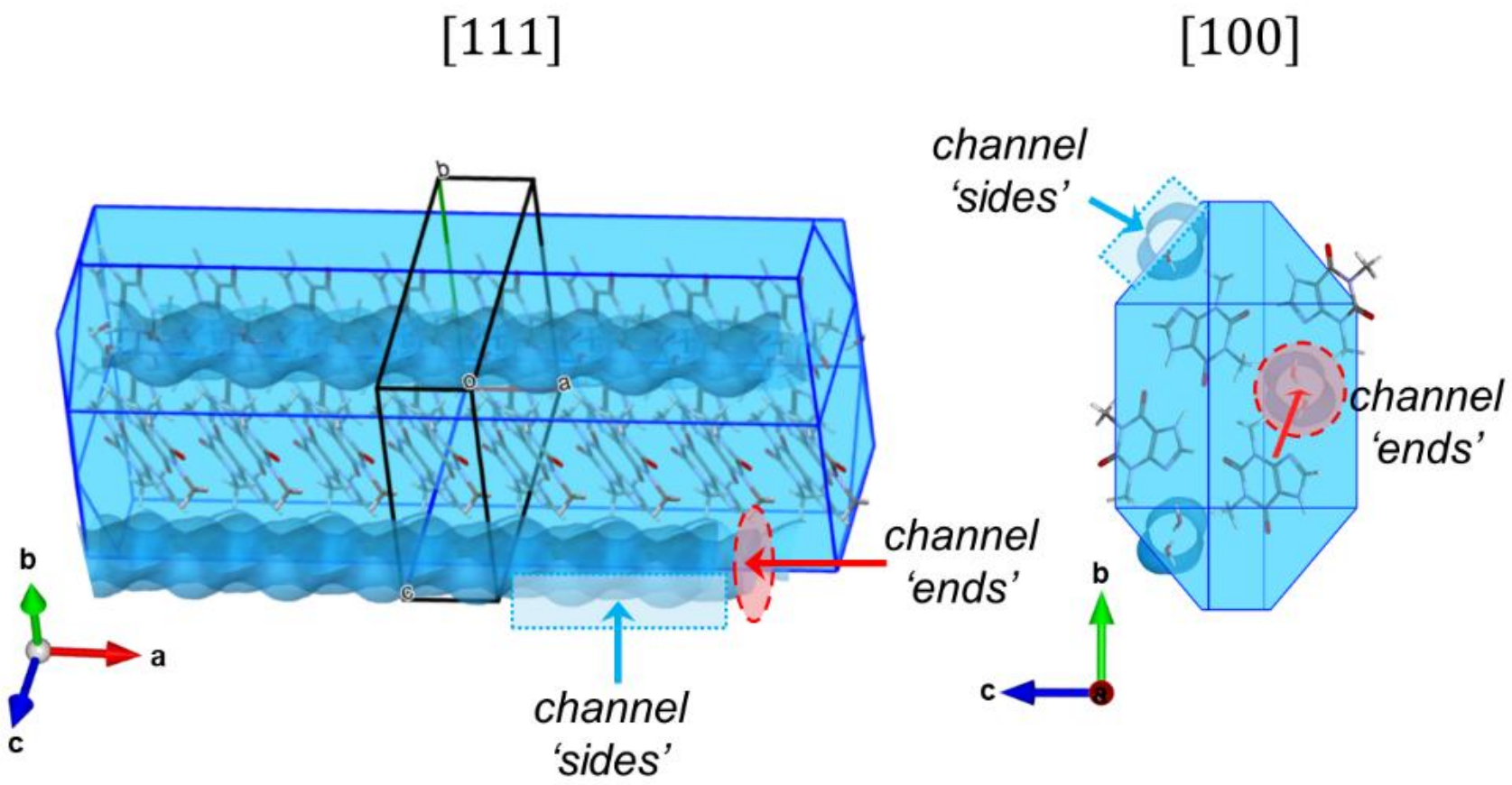

*Fig. S 3 BFDH morphology of theophylline monohydrate ($P2_1$) viewed along the [111] and [100] directions, illustrating the terminology used in this work. Channel 'sides' (blue arrows) refer to the long crystal surfaces parallel to the water channels, where the channel walls intersect specific surface facets, whereas channel 'ends' (red arrows) correspond to crystal faces perpendicular to the channel direction, where the water channels terminate.*

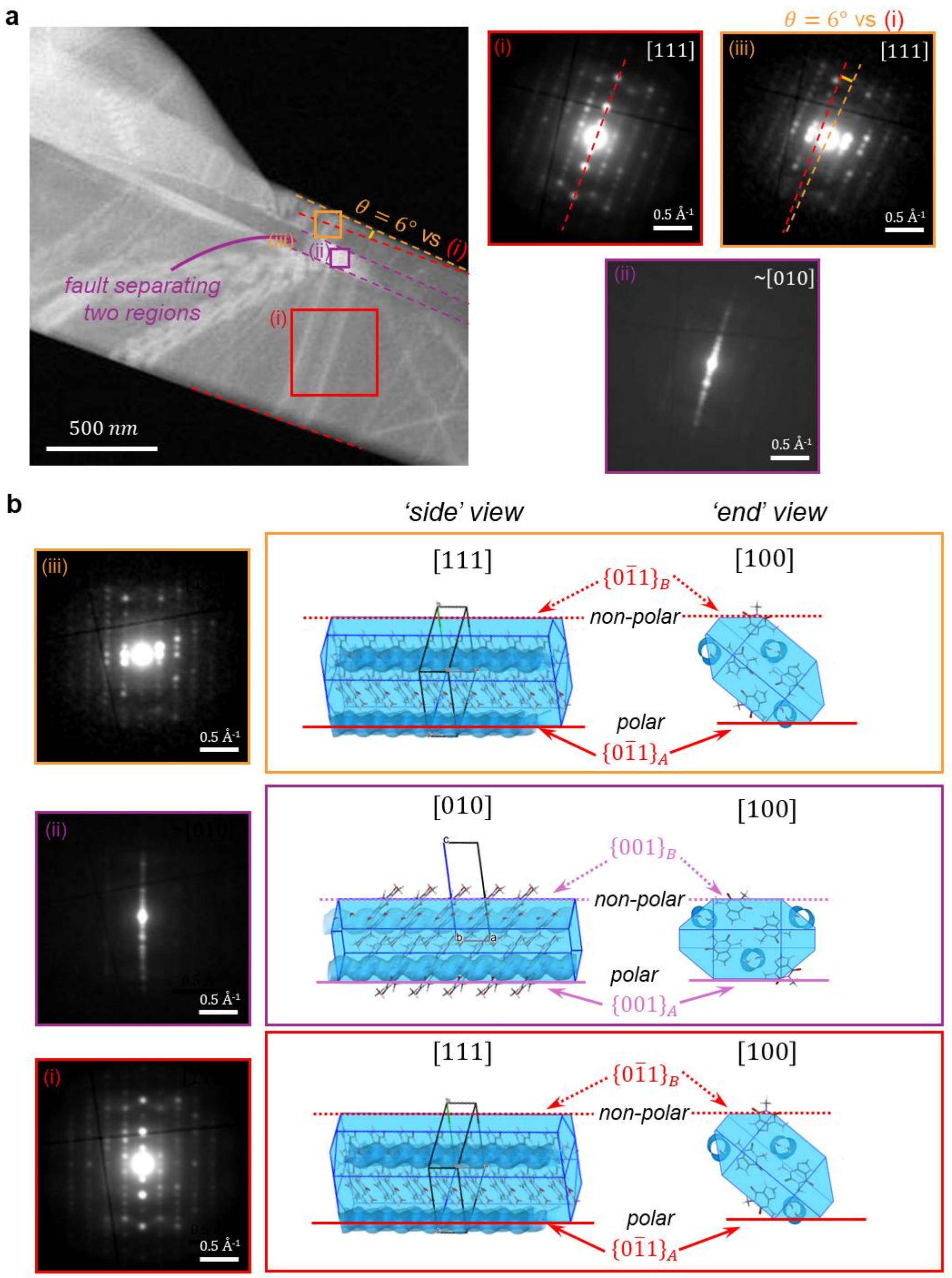

*Fig. S 4* ***a*** *Identification of three different crystalline areas within the particle in Fig. 3 and their relative orientation. Electron diffraction patterns taken from the upper and lower regions (labelled (i) and (iii) and highlighted by the red and orange squares, respectively) are both indexed to the* $[111]$ *zone axis, with a slight (~6°) in-plane rotation between corresponding diffraction vectors corresponding to* $\{hkl\} = \{0\bar{1}1\}$*. The inclined central fault (labelled by (ii) and marked by purple square), presumed to originate from a growth defect formed during crystallisation, is oriented along* $[010]$*. The geometry of the central fault produces the ~6° inclination between the upper and lower crystal regions, consistent with the corresponding shift of the main rows of reflections (marked by the dashed red and orange lines) in the diffraction patterns.* ***b*** *BFDH morphology models for each crystal orientation identified within the particle in* ***a****. For clarity, the corresponding diffraction patterns from* ***a*** *were rotated for simplified visualisation, such that the relevant plane directions are aligned vertically on the page. Accordingly, the models are aligned horizontally on the page for the 'side' view representation and oriented along [001] for the 'end' view representation. These models suggest the presence of water channels at the crystal interfaces, where the bottom interface of the* $[010]$ *oriented faulted region (ii) has the* $\{001\}_A$ *polar planes aligned to the* $\{0\bar{1}1\}_B$ *non-polar planes of the lower region (i) and vice-versa at the opposite, top interface.*

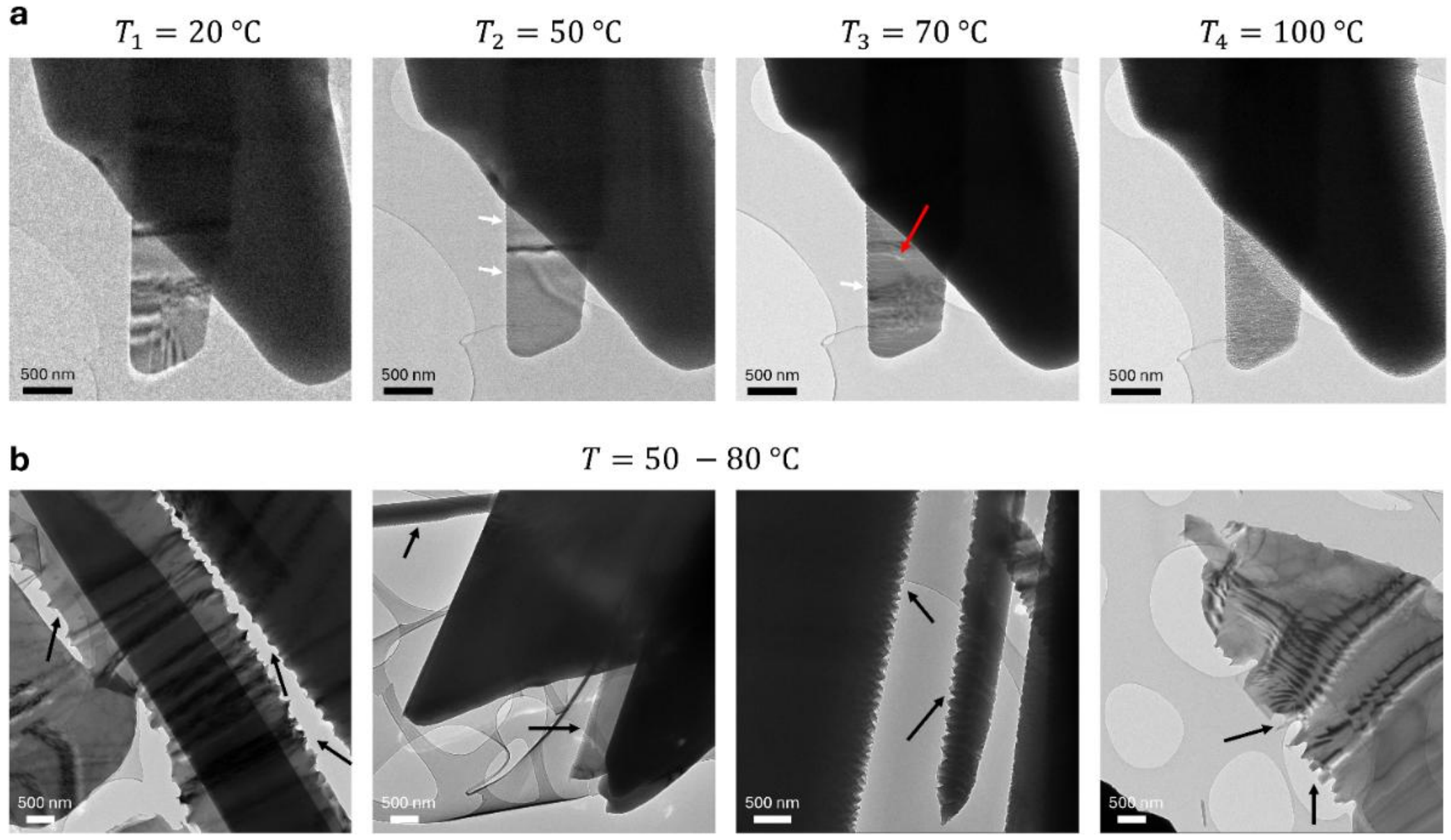

*Fig. S 5 In situ vacuum heating experiments in BF-TEM as an initial investigation of dehydration under condition 2.* ***a*** *Morphological changes of a representative crystal during heating from room temperature to 100 °C. Progressive surface roughening becomes visible at intermediate temperatures (marked by white arrows), leading to trench-like features (red arrow). Observations at the earliest stage of dehydration are consistent with condition 1.* ***b*** *Additional examples of different crystals imaged at intermediate temperatures (50 - 80 °C), showing increased surface roughening accompanied by mass loss and the onset of etching. Black arrows highlight regions exhibiting surface etching.*

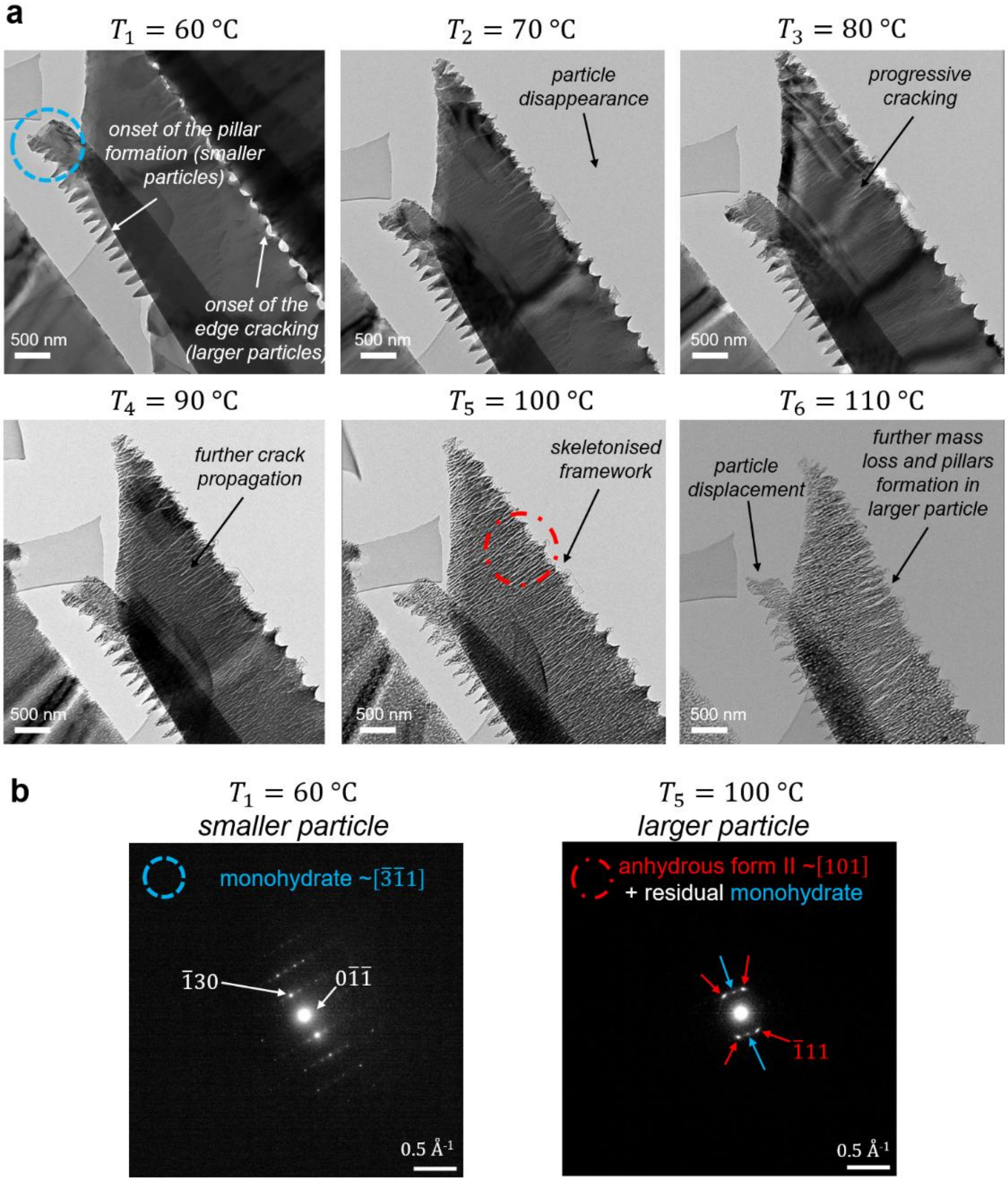

*Fig. S 6 Close-up analysis of the mid- to high-temperature regime during in situ vacuum heating experiments probing dehydration under condition 2. **a** Bright-field TEM images showing progressive morphological changes upon heating from 60 °C to 110 °C. A combination of thermal effects and electrostatic charging leads to multiple degradation features, including surface roughening, particle loss and displacement, mass loss, and the formation of well-defined pillar-like structures. Continued heating results in progressive mass loss and skeletal framework morphology. **b** Comparison of selected-area electron diffraction (SAED) patterns acquired from different regions and particle sizes (marked by blue and red circles). At 60 °C, diffraction patterns are consistent with the monohydrate phase. At 100 °C, reflections corresponding to the anhydrous form II are observed, while residual monohydrate reflections remain detectable, indicating phase coexistence. This spatial heterogeneity highlights the necessity of spatially resolved diffraction techniques, such as scanning electron diffraction (SED), to accurately resolve local dehydration pathways.*

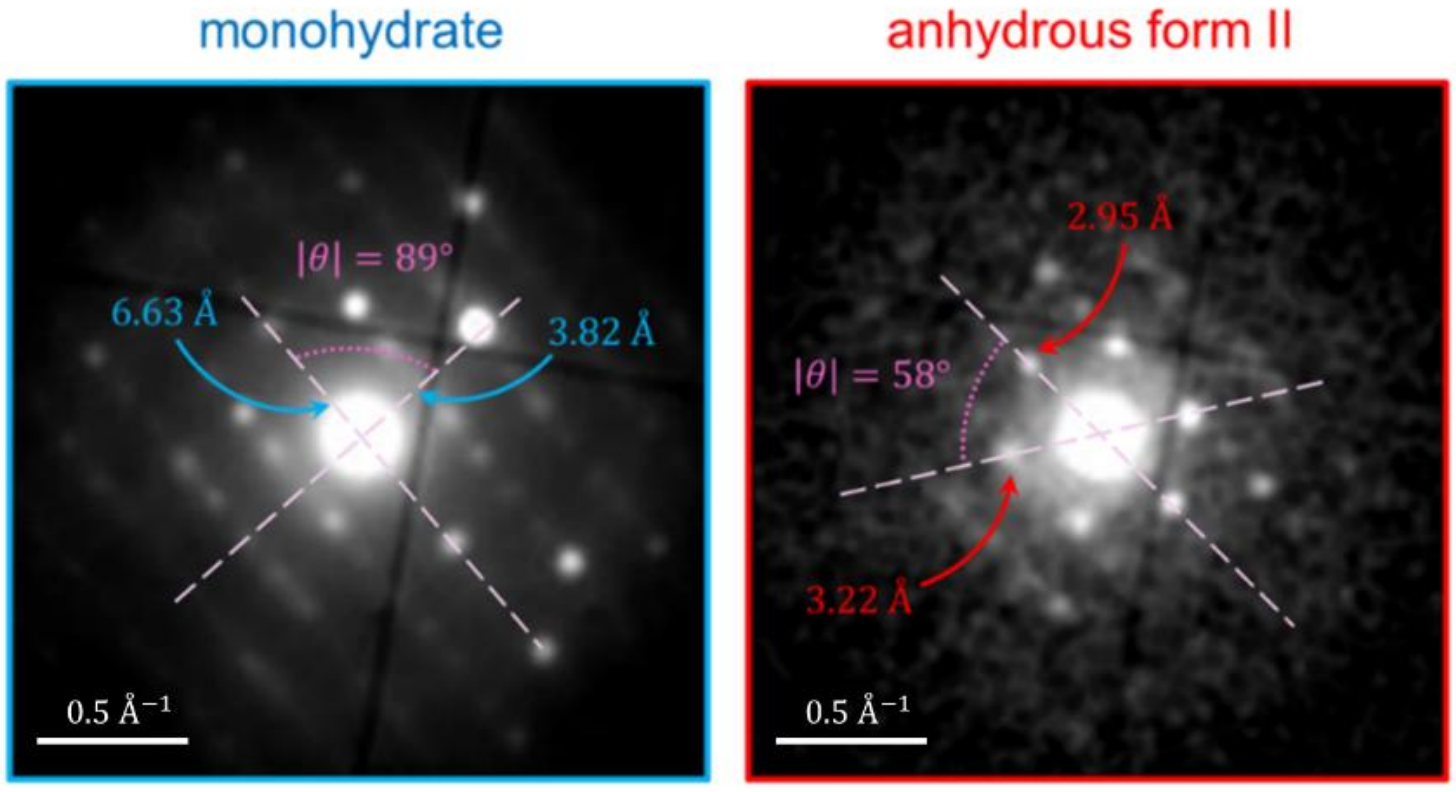

*Fig. S 7 Zoomed experimental diffraction patterns acquired under condition 2 (in situ vacuum heating). At 20 °C, the diffraction pattern is indexed to the theophylline monohydrate, while at 100 °C it is indexed to anhydrous form II. Reflections corresponding to the largest d-spacings (shortest reciprocal-space) and the relative angles between the associated lattice planes are indicated and used to determine possible zone axes and indexation.*

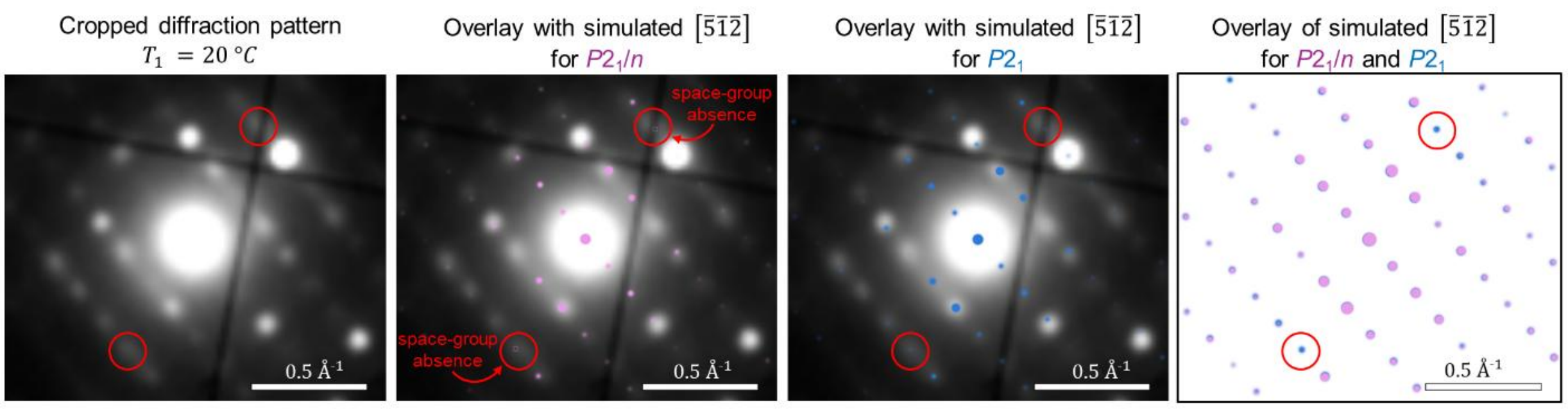

*Fig. S 8 Overlay of simulated electron diffraction patterns for theophylline monohydrate based on the $P2_1/n$ (purple) and $P2_1$ (blue) structures. The red circle highlights reflections forbidden by $P2_1/n$ space-group. Although these reflections are present in the experimental diffraction pattern, their intensity may arise from specimen thickness effects and double diffraction; therefore, it cannot be used conclusively to distinguish between the two space-groups listed here.*

*Table S 4 List of all evaluated zone-axis configurations between theophylline monohydrate and anhydrous form II, including relative rotations.*

| **Possible zone axis** | | **Possible rotation** |
|---|---|---|
| **monohydrate** | **form II** | |
| [$\bar{5}\bar{1}\bar{2}$]<br>[512]<br>[$\bar{5}\bar{2}\bar{1}$]<br>[521] | [$\bar{1}03$]<br>[$10\bar{3}$]<br>[103]<br>[$\bar{1}0\bar{3}$] | 0° and 180°<br>relative rotation between phase 2 vs 1 |
| *16 combinations* | | |
| *32 candidate configurations* | | |

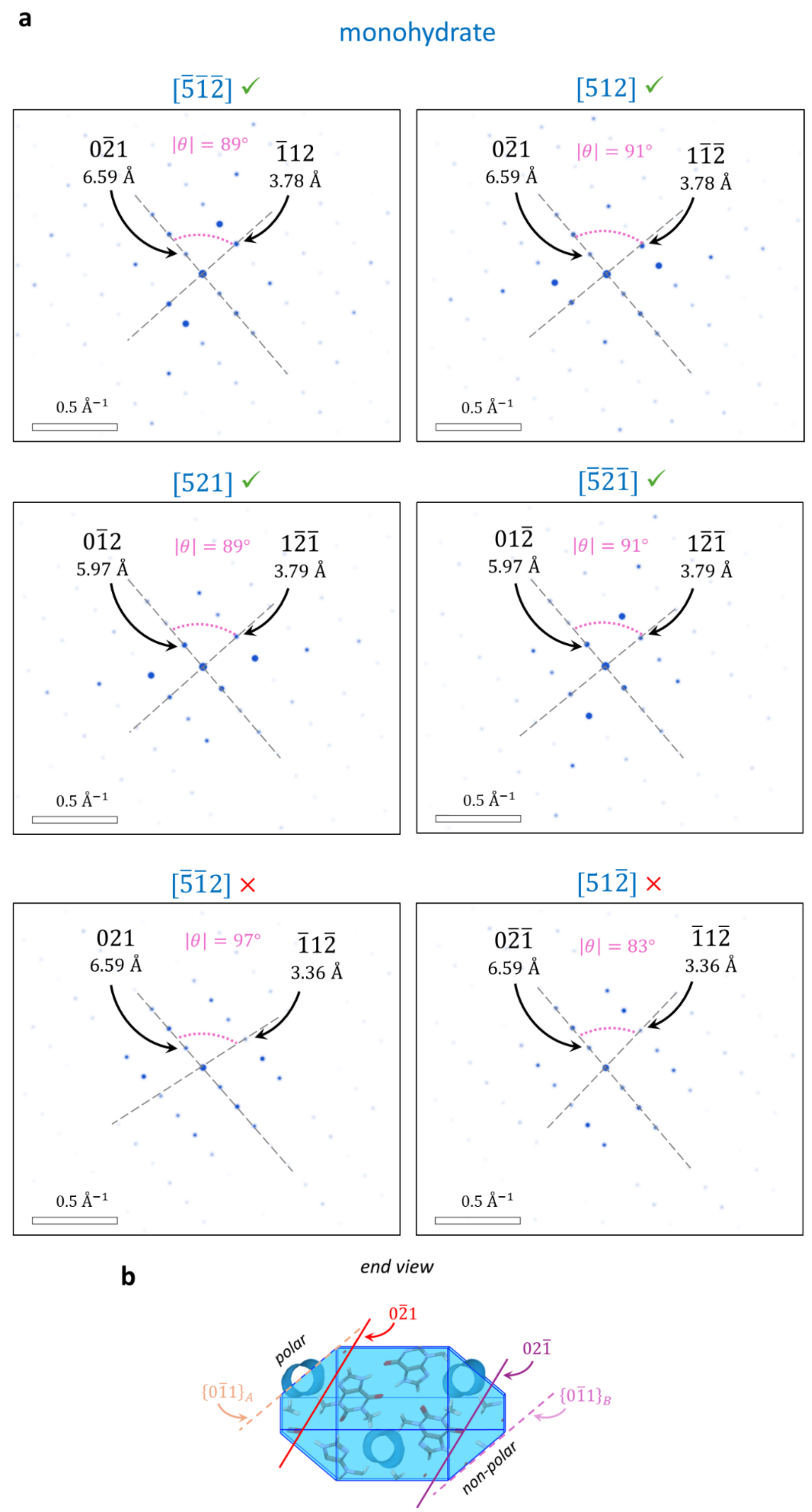

*Fig. S 9* ***a*** *Simulated diffraction patterns of theophylline monohydrate for candidate zone axes based on permutations of the monoclinic zone ⟨512⟩ and ⟨521⟩ that match the experimental diffraction pattern. The alternative permutation ⟨5̄1̄2⟩ is*

*excluded due to a mismatch in interplanar angles (the angle |θ| between the planes identified in the patterns is also reported in the patterns). Selected lattice planes are marked together with their corresponding d-spacings and the angles between these planes. b BFDH morphology model of P2₁ monohydrate crystal structure viewed [100] shows* $0\bar{2}1$ *is close to the polar* $\{0\bar{1}1\}$ *surfaces.*

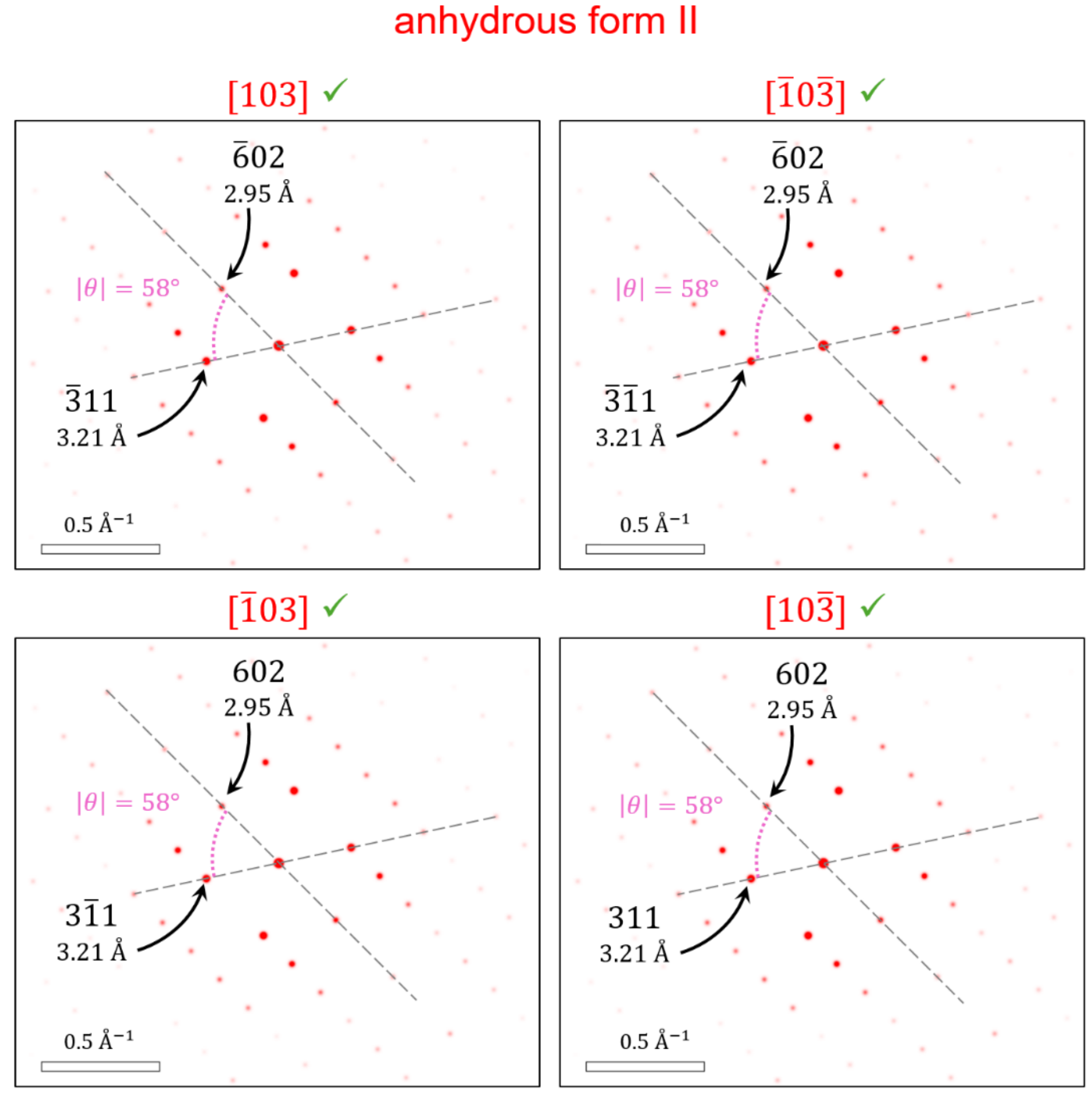

*Fig. S 10 Simulated diffraction patterns of anhydrous form II for candidate zone axes based on permutations of the orthorhombic zone* ⟨103⟩*. Selected lattice planes are marked together with their corresponding d-spacings and the angles between these planes.*

(i) Overlay of each combination of simulated diffraction pattern candidate pairs. A small shift in the spot positions of anhydrous form II is observed (towards smaller *d*-spacing) in the experimental diffraction pattern. The common row of reflections is highlighted in the overlay of the experimental diffraction patterns, and the original monohydrate reflections are marked by a blue line with dividers, while the shifted reflections corresponding to anhydrous form II are indicated by red arrows. Therefore, only candidate pairs that reproduce this shift in the anhydrous form II spot positions were considered.

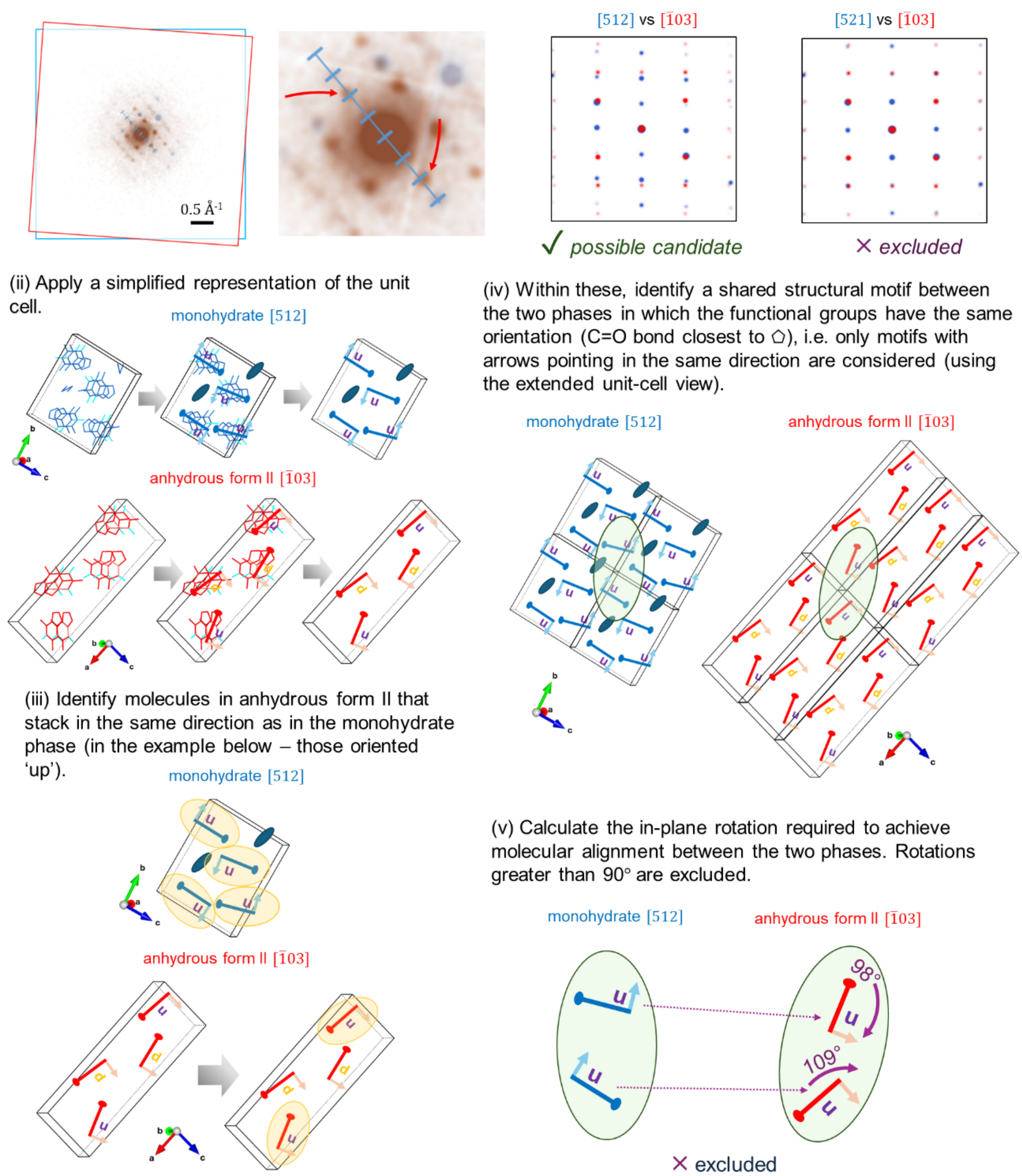

*Fig. S 11 Schematic of the screening procedure used to evaluate possible zone-axis configurations between theophylline monohydrate and anhydrous form II. The workflow illustrates the identification of possible zone axis candidate pairs, application of simplified geometric descriptors, identification of equivalently oriented molecular stacks, selection of shared structural motifs, and calculation of the required in-plane rotation.*

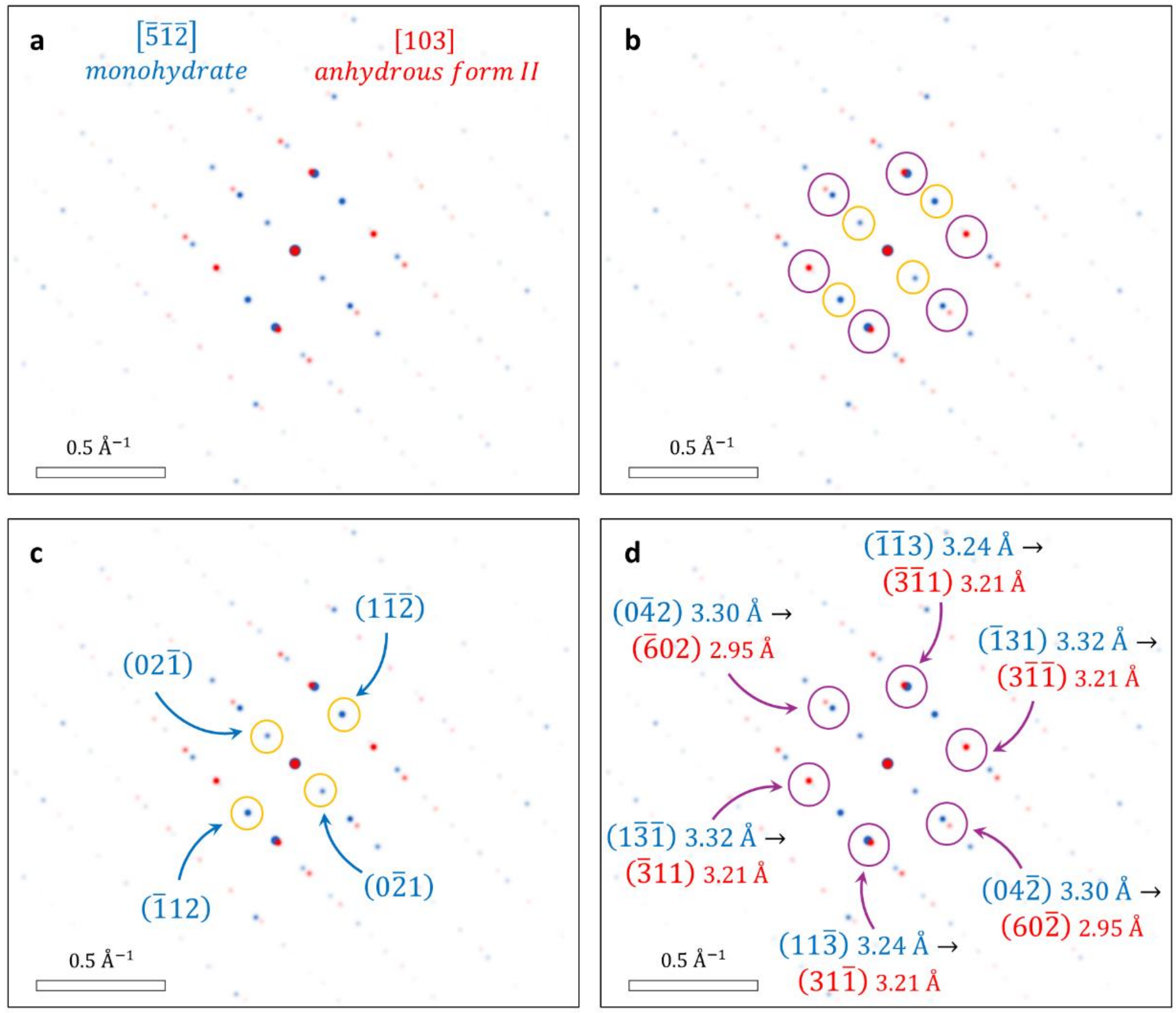

*Fig. S 12 Detailed analysis of simulated diffraction patterns of the theophylline monohydrate ($[\bar{5}\bar{1}\bar{2}]$, blue) and anhydrous form II ($[103]$, red). **a** Overlay of the simulated diffraction patterns after applying a 3° rotation to the pattern of anhydrous form II. **b** Reflections absent in the anhydrous form II are marked by yellow circles and are not visible after heating, while reflections that persist but are shifted are marked by purple circles. **c** Indexing of the monohydrate reflections marked by yellow circles. **d** Indexing of corresponding reflections present in both phases, together with their d-spacing values. The systematic shift of these reflections towards smaller d-spacings in anhydrous form II is consistent with lattice contraction upon dehydration.*

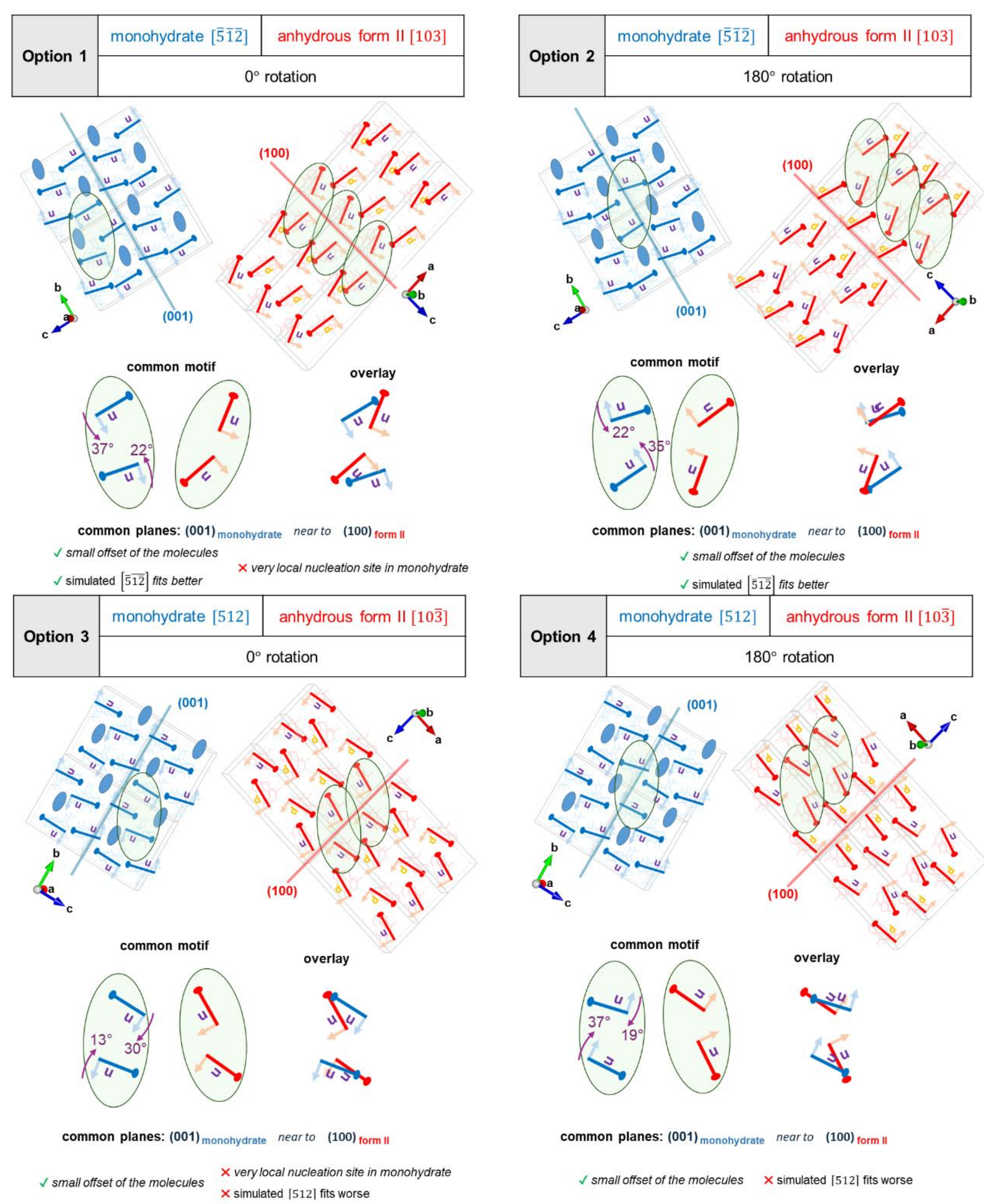

*Fig. S 13 Four possible zone-axis pairs between theophylline monohydrate and anhydrous form II showing preservation of a common molecular motif with small in-plane rotation. Option 2 was identified as the most plausible molecular pairing. Options 1 and 3 show registry limited to small nucleation regions within the monohydrate structure, while option 4 shows poorer agreement with the experimental diffraction patterns (*Fig. S 9*). However, all configurations identify the same common planes, namely (001) for the monohydrate and (100) for anhydrous form II.*

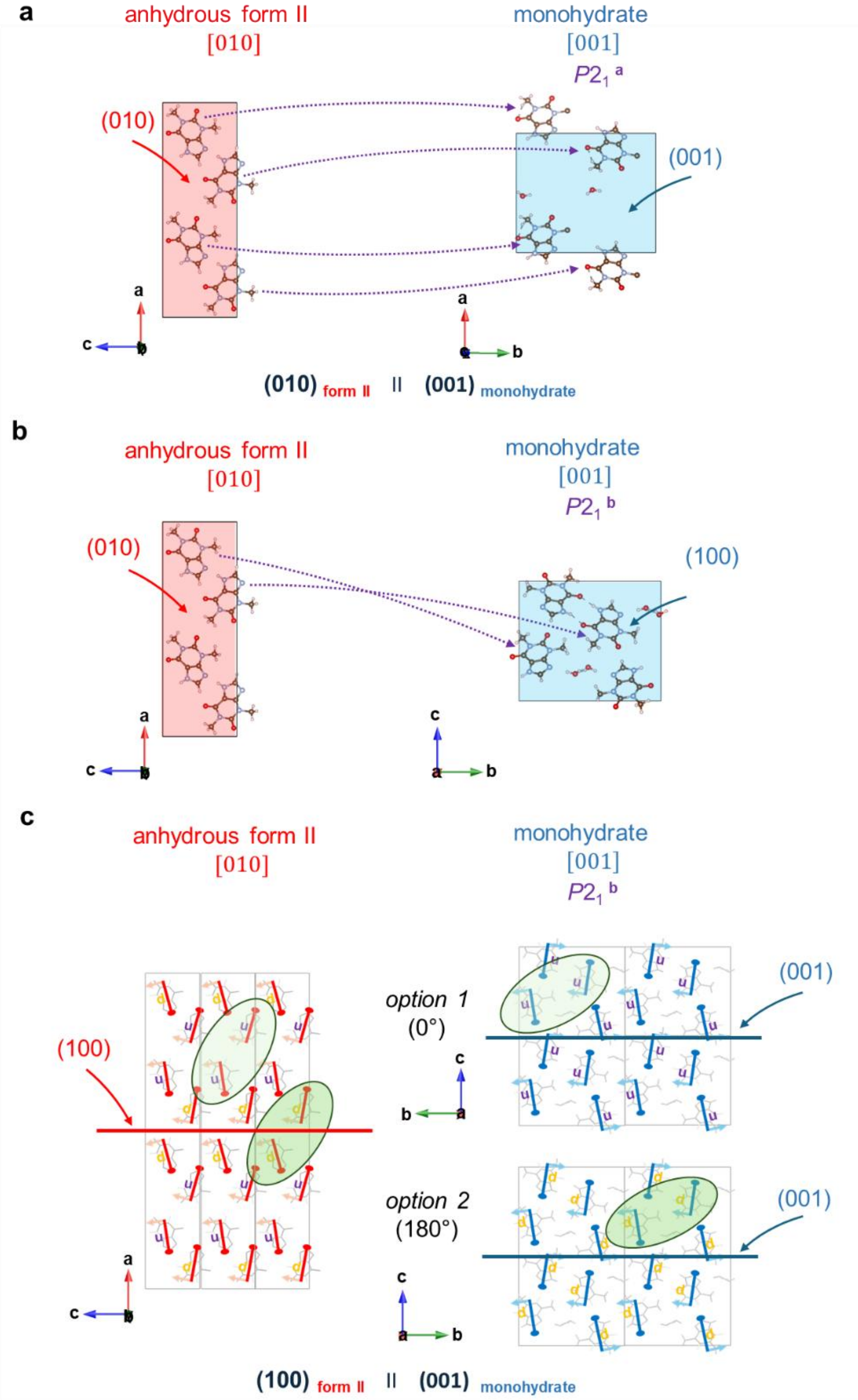

*Fig. S 14 Crystallographic reassessment of the reported epitaxial growth during hydration of theophylline. Rodríguez-Hornedo, N. et al. (41) inferred an epitaxial relationship between anhydrous form II and the monohydrate based on observations in which monohydrate needles grow on the face of anhydrous single crystals with a common plane of $(010)_{form\ II}$ II $(010)_{monohydrate}$. However, Rodríguez-Hornedo, N. et al. (41) report lattice parameters a, b and c as 13.3, 4.5 and 15.3 Å for the Sutor, D. (62) monohydrate structure in space group $P2_1{}^a$. In the CCDC entry (THEOPH), the same unit cell is given in an alternative setting with a, b and c as 13.3, 15.3 and 4.5 Å, corresponding to an interchange of the b and c axes. In this CCDC-consistent setting, the proposed orientation relationship is $(010)_{form\ II}$ II $(001)_{monohydrate}$. **a** A schematic showing the*

*common plane proposed by Rodríguez-Hornedo, N. et al. (41) in which four molecules match across the two phases. We show this plane in the $P2_1{}^a$ structure of the monohydrate (62).* ***b*** *Re-evaluation using the corrected $P2_1{}^b$ monohydrate structure (49), rather than the earlier $P2_1{}^a$ crystal structure (shown to be incorrect), shows that, on the originally inferred common plane between anhydrous form II and the monohydrate, only two molecules exhibit similar in-plane geometry (however, the direction of the molecular stacking is not considered here).* ***c*** *Simplified geometric modelling was used to identify plausible common motifs between the anhydrous and corrected $P2_1{}^b$ crystal structure. Candidate common planes were identified as $(100)$ in the anhydrous form II and $(001)$ in the monohydrate. The monohydrate $(001)$ plane is consistent with the plane identified in our experimental in situ dehydration experiment. This reassessment provides a structurally consistent interpretation linking the previously reported hydration orientation relationship with the low-index planes observed during dehydration, indicating that either process involves crystallographically related interfacial motifs.*

*Fig. S 15 Proposed two-step dehydration model of theophylline monohydrate involving (i) surface dehydration close to $\{0\bar{1}1\}_A$ polar planes with pillar formation and (ii) templated crystal growth of anhydrous form II. Representative ADF images, diffraction patterns, and unit-cell schematics corresponding to the indexed orientations are shown to visualise the proposed transformation mechanism.*